# Revealing the Physico-Chemical Basis of Organic Solid-Solid Wetting Deposition: Casimir-Like Forces, Hydrophobic Collapse, and the Role of the Zeta Potential

Alexander Eberle,[†,‡] Thomas Markert,[§] and Frank Trixler*,[†,∥]

[†]Department of Earth and Environmental Sciences and Center for NanoScience (CeNS), Ludwig-Maximilians-Universität München, Theresienstraße 41, 80333 München, Germany

[‡]Helmholtz Institute Erlangen-Nürnberg for Renewable Energy, Egerlandstr. 3, 91058 Erlangen, Germany

[§]Institute of Theoretical Chemistry, Ulm University, Albert-Einstein-Allee 11, 89081 Ulm, Germany

[∥]TUM School of Education, Technical University of Munich and Deutsches Museum, Museumsinsel 1, 80538 München, Germany

**ABSTRACT:** Supramolecular self-assembly at the solid-solid interface enables the deposition and monolayer formation of insoluble organic semiconductors under ambient conditions. The underlying process, termed as the Organic Solid-Solid Wetting Deposition (OSWD), generates two-dimensional adsorbates directly from dispersed three-dimensional organic crystals. This straightforward process has important implications in various fields of research and technology, such as in the domains of low-dimensional crystal engineering, the chemical doping and band-gap engineering of graphene, and in the area of field-effect transistor fabrication. However, till date, lack of an in-depth understanding of the physico-chemical basis of the OSWD prevented the identification of important parameters, essential to achieve a better control of the growth of monolayers and supramolecular assemblies with defined structures, sizes, and coverage areas. Here we propose a detailed model for the OSWD, derived from experimental and theoretical results that have been acquired by using the organic semiconductor quinacridone as an example system. The model reveals the vital role of the zeta potential and includes Casimir-like fluctuation-induced forces and the effect of dewetting in hydrophobic nano-confinements. Based on our results, the OSWD of insoluble organic molecules can hence be applied to environmental friendly and low-cost dispersing agents, such as water. In addition, the model substantially enhances the ability to control the OSWD in terms of adsorbate structure and substrate coverage.

## INTRODUCTION

Supramolecular self-assembly, utilizing programmable, non-covalent interactions, enables the bottom-up fabrication of low-dimensional nanostructures and adsorbates using organic semiconductor molecules, for applications such as carbon-based nanoelectronics[1-7] and crystal engineering.[8-12] There are two common technologies to perform the corresponding bottom-up assembly: vapor deposition and liquid phase deposition techniques.[13,14] However, both the approaches possess distinct limitations: the vapor deposition methods, like the organic molecular beam deposition,[14,15] being only applicable to few organic substances that survive a thermally enforced vacuum sublimation.[16-20] Liquid phase deposition techniques, as drop-casting or spin-coating,[16] are based on chemical solutions, thus being unable to incorporate most of the organic pigments with promising semiconductive properties, owing to their insolubility in almost all of the liquid media. The latter drawback hence limits the processing of the organic pigments without a chemical functionalization that otherwise enables their dissolution.[6,20] However, the customized synthesis of functionalized semiconductors is expensive, particularly in relation to the standard pigments already being used in the industry.

Hence, due to the above limitations, an alternative deposition approach was developed, termed as the "Organic Solid-Solid Wetting Deposition" (OSWD).[17-21] This new deposition approach possesses several advantages, as being environmental friendly, cheap, and both straightforward and quick to perform under ambient conditions. To induce the OSWD, typically a powdered organic semiconductor such as quinacridone, acridone or perylenetetracarboxylic dianhydride (PTCDA) is dispersed in a dispersing agent and then drop-casted on a substrate, such as graphite, graphene, carbon nanotubes or MoS$_2$.[18-20,25] Subsequently, two-dimensional (2D) adsorbates are formed directly from the three-dimensional (3D) particles, as the dispersed particles get in touch with the substrate. The deposition is driven by the solid-solid wetting effect,[22-24] with a gradient of the surface free energy serving as the driving force. Applications of the OSWD have become relevant e.g. in the fields of low-dimensional crystal engineering on surfaces,[21,26-27] the chemical doping and bandgap engineering of graphene,[19] and in the fabrication of organic field-effect transistors.[16]

However, until now, lack of profound understanding of the physico-chemical basis of the OSWD has restricted this technology to a narrow range of suitable dispersing agents. This limitation prevented the broader development of ways to catalyze the basic effect, and hence, excluded its possible applications in (bio)nanotechnology. Thus, the aim of the current study was to provide a detailed model of the OSWD process by identifying the contributing physical and chemical forces and the required environmental conditions. For this, an example system built by a highly oriented pyrolytic graphite (HOPG) as the substrate and dispersed crystalline particles of the organic semiconductor gamma quinacridone (γQAC) as the adsorptive, was employed. γQAC was chosen in this regard owing to its absolute insolubility in water[35-38] and in common organic solvents at atmospheric pressure,[30,38] making it thus highly suitable for solid-solid wetting studies. Further, among the various QAC polymorphs that have been identified so far, γQAC was found to be the most stable form,[38] possessing further promising electrical properties, low toxicity, excellent physical and chemical stability,[28-34] and being easily available as a commercial pigment. With the aid of the example system, the OSWD process was explored *ceteris paribus* by analyzing various dispersions in terms of particle size, zeta potential, pH, and by imaging QAC adsorbate structures via scanning tunneling microscopy (STM). Additional insights were obtained via force field simulations and calculations based on a refined DLVO theory that integrates fluctuation-induced forces.

## THE SPECIFICS OF THE OSWD

**How can the OSWD be induced?** In a standard OSWD preparation procedure, a fine powder of organic molecular crystals (for instance, organic pigments, organic semiconductors) is mixed with a liquid (as organic liquids, water), with the powder share representing 2 - 10 weight %. The insoluble molecular crystals form a particle dispersion in the mixture, thereby the liquid acting as the dispersing agent. These dispersed organic crystals are next drop-casted under ambient conditions onto an inorganic substrate, such as graphite. Depending on the chosen dispersing agent, the result of the OSWD can be probed via STM under ambient conditions either directly by scanning within the dispersion, or post the drop-casted dispersion is dried and the dry sample surface is subsequently covered with a thin layer of a liquid alkane. For the latter, a liquid alkane was found to neither trigger OSWD nor form supramolecular assemblies, it just has the function of preventing the formation of a contamination layer, condensed from the air.[39-40] Further, though the preparation procedure is very straightforward in itself, understanding the OSWD formation process and the outcomes of the technique is very challenging and was the aim of this study.

**What is the result of the OSWD?** OSWD results in the formation of 2D adsorbate monolayers directly from insoluble but dispersed 3D particles.[17-19] Regarding our example system, QAC molecules (Figure 1) form one-dimensional (1D) supramolecular chains via NH···O=C hydrogen bonding. The supramolecular chains have a uniform width of 1.63 nm and can arrange into multiple parallel and side-by-side appearing arrays (Figure 2).[18] In large-scale STM scans, these arrays are imaged as linear features and domains, as can be seen in the Figure 4. Further, QAC bilayer structures were found occasionally, indicating their detectability but absence in most of the images (refer supporting information for related example images).[18]

**What is the role of the dispersing agent?** Over a period of several weeks, the simple coverage of the HOPG substrate with dry γQAC powder does not lead to adsorbate structure formation, according to our STM analysis. This result corresponds to previous findings, showing that the OSWD cannot be trig-gered even by heating up γQAC powder atop a HOPG to a temperature of 160 °C.[19] In addition, experiments were performed as to generate QAC adsorbate structures by grinding γQAC powder on a HOPG using a metal spatula. Results revealed that soft grinding does not lead to the formation of supramolecular QAC structures on the HOPG surface, whereas a harsher grinding damages the HOPG surface, making STM analysis thus impossible. In summary, these results hence indicate that at ambient conditions, the OSWD cannot be triggered without a dispersing agent.

Now the question that arises here is: why is the presence of a dispersing agent so vital for an OSWD to occur under ambient conditions? To answer this, the impact of different dispersing agents on the performance of the OSWD process was studied, as to analyze the respective surface coverage with adsorbate arrays. For this, a series of tests were initiated, using both organic liquids and purified water as the dispersing agents. Results revealed that the median of the overall surface coverage of the HOPG by QAC adsorbate structures (including both single appearing 1D QAC chains and 2D QAC arrays) differs appreciably, depending on the dispersing agent in use (Figure 3). The lowest median of the surface coverage rate was found when dodecane was employed (0.3 ± 0.3 %, including twice the standard deviation), whereas the highest coverage of 75 ± 19 % was achieved for the propylene carbonate case. The obtained results hence suggest that the role of the dispersing agent is to catalyze the OSWD process, for it to be occurring under ambient conditions.

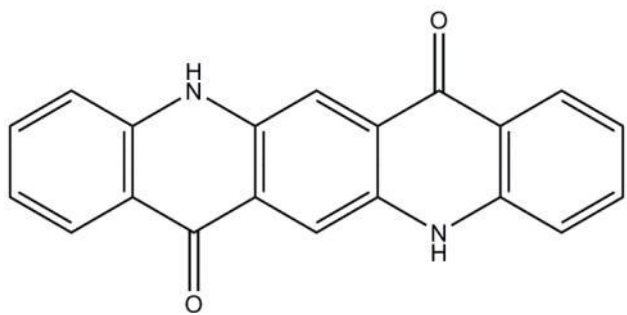

**Figure 1.** Chemical structure of the QAC molecule.

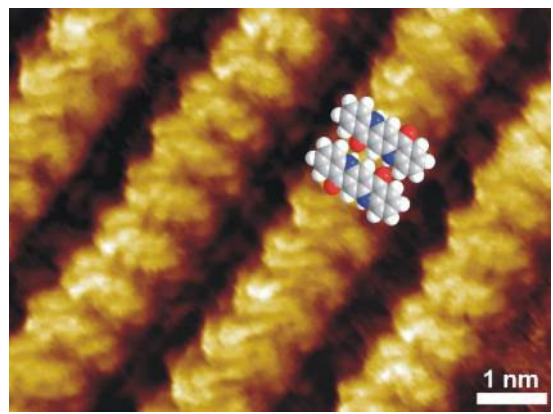

**Figure 2.** STM close-up view of an QAC adsorbate layer, generated via the OSWD. Additionally, an inlay showing the force field simulated arrangement of two QAC molecules highlights, how supramolecular chains are formed via NH···O=C hydrogen bonds.

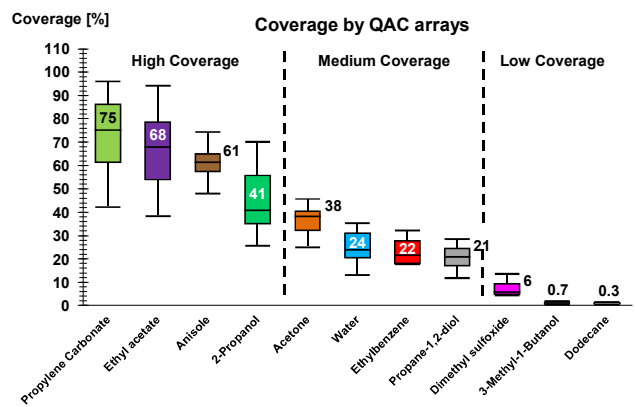

**Figure 3.** Box plot diagram of the coverage of the HOPG surface by QAC adsorbates, corresponding to the dispersing agents used for the sample preparation. With the specified values refering to the median of the surface coverage, the different samples have been roughly classified into three classes, as to compare the effectivity of the OSWD in a straightforward way: the "High Coverage", comprising samples with an average coverage of $100 - 40$ %; the "Medium Coverage", comprising samples with an average coverage of $40 - 15$ %; and the "Low Coverage", incorporating samples with no QAC arrays but single occurring QAC chains with an average coverage of $15 - 0$ %.

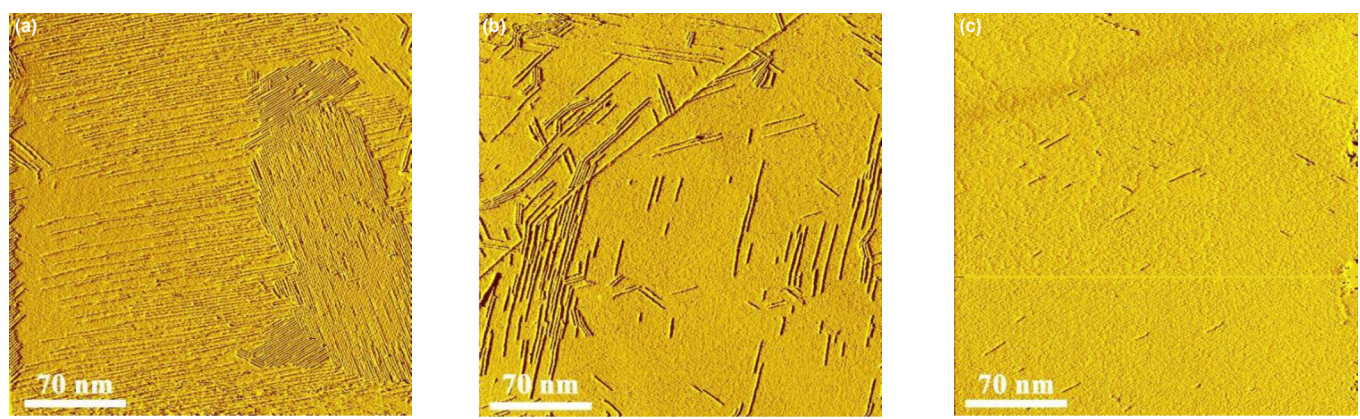

**Figure 4.** STM images showing exemplary the HOPG surface coverage (SC) by QAC adsorbates that have been generated using different dispersing agents. The surface coverage classified as the "High Coverage": (a) Ethyl acetate, SC = $65 \pm 15$ %; the "Medium Coverage": (b): Propane-1,2-diol, SC = $20 \pm 5$ %; and the "Low Coverage": (c) 3-Methyl-1-Butanol, SC = $0.7 \pm 0.7$ %.

## EXPERIMENTAL IDENTIFICATION OF PARAMETERS INFLUENCING OSWD

The results obtained so far suggest, that the quantity of supramolecular adsorbate structures generated via OSWD is related to the properties and conditions of the catalyzing dispersing agent in use. To further investigate this finding, efforts were made to correlate different physical properties of the used dispersing agents to the median of the achieved surface coverage rate. The coefficient of determination $R^2$ (square of the Pearson correlation coefficient R) was used in this regard, to quantify the degree of a potential (linear) correlation (refer supporting information for more details on the related scatter plots).

**Which parameters do not influence the OSWD?** The results of our investigations yield no correlation between the surface coverage and the viscosity ($R^2 = 3$ %), the permittivity ($R^2 = 0.7$ %), the surface tension ($R^2 = 2$ %), or the vapor pressure ($R^2 = 6$ %) of the investigated dispersing agents. However, worth examining her is: could the dispersing agents in any way cause specific modifications to the properties and conditions of the dispersed $\gamma$QAC crystals? It is a well-known fact that the particle size distribution of a dry powder alters when dispersed, owing to agglomeration and anti-agglomeration processes.[42]

Additionally, the size of a nanocrystal influences its surface free energy,[41] indicating a potential relationship between the crystal size and the effectivity of the OSWD process. However, a corresponding analysis yielded $R^2 = 34$ % for a correlation between median particle size and surface coverage, indicating thereby no significant impact of the particle size on the OSWD process (refer supporting information for figures depicting the particle size analysis).

**Which parameters influence an OSWD?** An important consequence of the dispersing procedure is the generation of surface charges that induce a zeta potential. The zeta potential is generally known to significantly determine the interdependency of dispersed particles, irrespective of whether the dispersing agent is a polar or non-polar liquid.[42,43] Additionally, Zhao et al. could demonstrate that the surface free energy of a solid is modified via surface charges,[48] indicating a possible relationship between the zeta potential and the solid-solid wetting processes. Measurements in this regard revealed that each dispersing agent generates its individual zeta potential distribution (refer the supporting information for a corresponding diagram depicting all the determined zeta potential distributions).

It has further been shown by Zhao et al., that the surface free energy of a solid drops with an increasing surface charge,[48] suggesting that triggering an OSWD process may be promoted by dispersed γQAC particles, exhibiting a high zeta potential. Based on this, key indicators of the determined zeta potential distributions, representing the strongly charged γQAC fractions, have been examined for a potential correlation. As a result, a significant correlation ($R^2$ = 74 %) was found between the surface coverage and the z33 value, referring to the point where 33 % of the zeta potential distribution is more negative and 66 % is more positive (Figure 5). The result indicates a significant influence of the zeta potential on the OSWD process, as it can be assumed that the OSWD process is governed by a complex balance of interactions between various components in the system, making the determination of a linear independence of the achieved surface coverage and a physical parameter implausible. It was thus decided, to further explore the potential of directing and fine-tuning the OSWD process, via modifying the zeta potential.

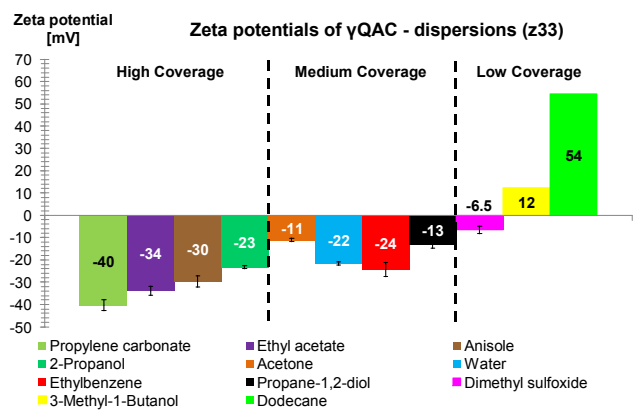

**Figure 5.** Comparison of the zeta potential values z33 (i.e. the point where 33 % of the distribution is more negative and 66 % is more positive), derived from γQAC dispersions with varying dispersing agents. The error bars indicate twice the standard deviation. Please note that the measuring method determining the zeta potential distribution failed in dodecane and in 3-methyl-1-butanol. The indicated zeta potential values were thus determined by an alternative measuring method yielding an average value that demonstrates the γQAC crystals in dodecane and in 3-methyl-1-butanol to be conversely charged. However, to guarantee comparability, these average values were not incorporated in the correlation analysis.

## FINE-TUNING ZETA POTENTIALS IN AQUEOUS DISPERSIONS, TO UNDERSTAND BETTER THE OSWD

The well-established concept describing the origin of the zeta potential is based on the model of the electrical double layer.[44] A decisive role within this model play dissociated salts (i.e. ions), as they modify the surface charge of dispersed particles and the extent of the diffuse layer. Thus, in order to fine-tune zeta potentials in a *ceteris paribus* approach, as to obtain a deeper understanding of the OSWD process, modification of the zeta potential with the aid of different salts was attempted.

In this regard, phosphate salts, like sodium triphosphate, disodium phosphate, and disodium pyrophosphate are known to stabilize aqueous dispersions by pushing the zeta potential to a higher electronegative region.[46] Tests were additionally performed using sodium chloride, the most common salt worldwide, and adenosine 5'-monophosphate disodium salt, serving as a main source of energy in biological cells.[45]

Analysis revealed that the addition of salts to the aqueous γQAC dispersions led to modification of the z33 value in the range of - 4 to - 45 mV (Figure 6 (a)), resulting thus in the median of the surface coverage (SC) of the HOPG to be between 12 and 43 % (Figure 6 (b)). Three salts that increased the coverage in this respect were: adenosine 5'-monophosphate disodium (SC = 43 %), disodium phosphate (SC = 41 %), and sodium triphosphate (SC = 30 %). Further, the addition of disodium pyrophosphate (SC = 17 %) and sodium chloride (SC = 12 %) had a tendency towards constraining the assembly of the QAC arrays.

As the zeta potential is also related to the pH for obvious reasons,[46-47,50] tests were thus performed towards exploring the potential influence of the pH value on the OSWD process. It was found that the addition of salts modifies the pH value (Figure 7) and further analysis revealed a tendency towards a dispersion with a high pH inducing a high surface coverage (Figure 6 (b)). Additional tests revealed that increasing the pH by adding a base (KOH) shifts the zeta potential to a more negative side (Figure 6 (a)) and increases the resulting surface coverage (Figure 6 (b)), whereas such an effect was observed to be reversed by the addition of an acid (e.g. $H_2SO_4$). Further, the co-efficients of determination for the pH ($R^2$ = 55 %) and the z33 of the aqueous samples ($R^2$ = 58 %), show evidence of a correlation of these parameters to the surface coverage as well. To evaluate the potential influence of the substrate's surface charge (generated as soon as the HOPG is wetted by a dispersion) on the OSWD process, additional measurements were performed, as to analyze the zeta potential of graphite particles dispersed within the hitherto analyzed aqueous systems. Results revealed no significant correlation ($R^2$ = 26 %) of the z33 of the HOPG to the achieved surface coverage (Figure 8).

However, a question that arises here is: can these results be influenced by chemical interactions related to the performed pH modifications or the addition of a salt? Within the pH range of 1 – 12, QAC molecules were found to be chemically stable and the crystal structure of QAC polymorphs was found to be unaffected by pH modifications.[49] Further, experimental results revealed that $Na^+$, $Cl^-$, and $SO_4^{2-}$ ions, added to aqueous systems in concentrations of ≥ 0.1 mol l⁻¹, do not react chemically with QAC (the salt concentrations used within this study were below 0.0043 mol l⁻¹).[49] Thus, for an OSWD process, the participation of pH related chemical processes or chemical reactions between the added salts and QAC can be excluded.

In addition, with γQAC being both chemically inert and strongly hydrophobic,[35-38] the ionization of surface groups and a related strong affinity for ions can be excluded.[50] Moreover, several studies yielded no specific adsorption of ions by hydrophobic surfaces below a salt concentration of 0.01 mol l⁻¹, except for the adsorption of hydroxyl ions.[50] Experimental investigations and molecular dynamics simulations in this regard revealed that the water molecules form an oriented ice-like structure at extended hydrophobic surfaces, as a result of the competition between the tendencies of the liquid to maximize the number of hydrogen bonds and to maximize the packing density.[51] The water structuring makes hydrogen bonds between hydroxyl ions and water molecules energetically favorable, leading to the preferential adsorption of hydroxyl ions and their subsequent stabilization.[50-51] Accordingly, the modification of the zeta potential by the addition of salt is solely related to the simultaneously occurring modification of the pH, but is not related to the adsorption of salt ions, provided that the salt concentration remains low.[50] Thus, referring to the low salt concentrations used within this study, we conclude that supramolecular interactions between the added salts and the surface molecules of the dispersed $\gamma$QAC particles do not have a significant impact on the via OSWD generated surface coverage.

Summarizing the results obtained so far, it can hence be said that under ambient working conditions, triggering of the OSWD requires a catalyzing dispersing agent. Outcomes further indicate that the gradient of the surface free energy between an organic semiconductor and the substrate surface is related to the zeta potential. Besides, fine-tuning the zeta potential and the related pH via salts, acid or base added to aqueous dispersions enables the control of the OSWD in terms of the surface coverage, opening the way to a deeper understanding of the OSWD process via theoretical approaches.

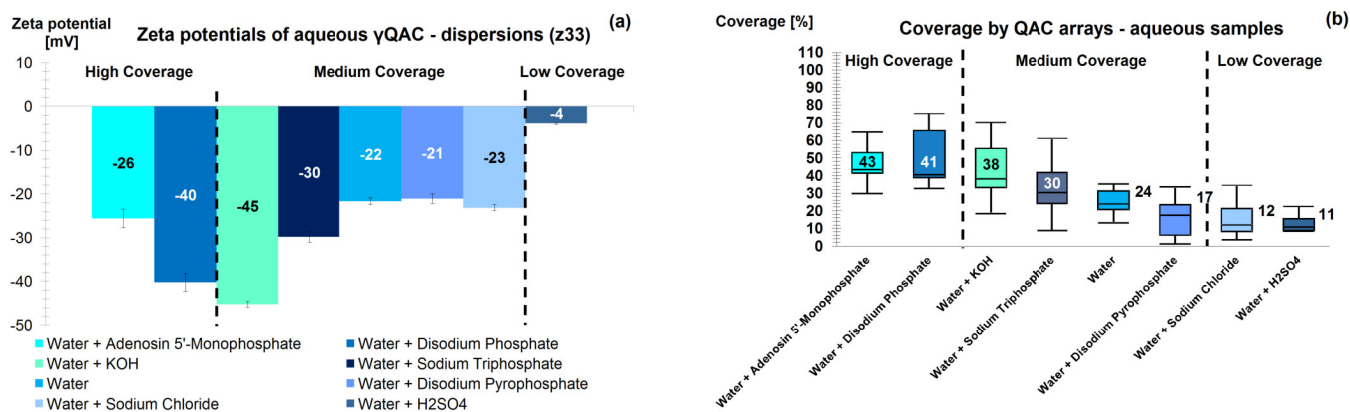

**Figure 6.** (a) Depicting of the zeta potentials (z33) of aqueous $\gamma$QAC dispersions modified via different salts, the acid $H_2SO_4$, and the base KOH. (b) The corresponding median values of the HOPG surface coverage rates.

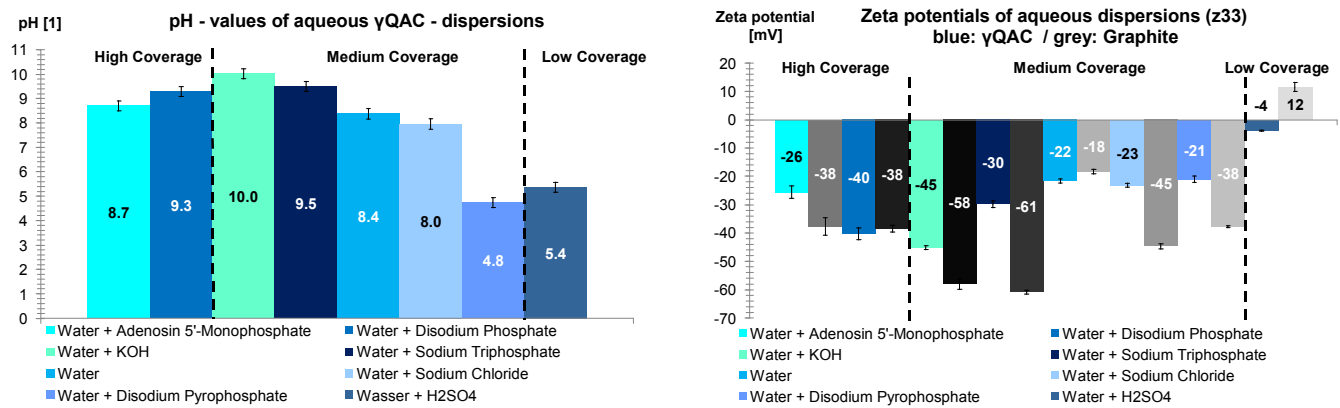

**Figure 7.** Comparison of the the pH values of the analyzed aqueous $\gamma$QAC dispersions, treated with either salts, $H_2SO_4$ or KOH.

**Figure 8.** Comparison of the zeta potentials (z33) of dispersed $\gamma$QAC crystals and graphite powder (equivalent to the z33 of the HOPG surface) in all the tested aqueous systems.

## THEORETICAL APPROACHES TOWARDS UNDERSTANDING THE OSWD

**DLVO calculations.** To further explore this potential interdependency among the OSWD surface coverage, the gradient of the surface free energy (between an organic semiconductor and the substrate), the related zeta potential, and the pH, an appropriate mathematical model was used. This was done as to simulate the approach of a charged semiconductor crystal ($\gamma$QAC) towards the substrate (HOPG) surface, and to approximate the related gradient of the surface free energy. In this regard, it was hardly possible to calculate the local surface free energy, especially when the special case of a solid-solid wetting process is involved. However, the DLVO theory, in this respect, presented itself as a suitable model, enabling the calculation of the potential energy of the interaction between two particles by simply summing up the attractive (i.e. Van-der-Waals) and the repulsive (i.e. double layer or Poisson-Boltzmann, respectively) components, as a function of their separation.[42]

Further, though the traditional DLVO model has well demonstrated its validity in describing colloidal dispersions,[43] it comprises solely of the repulsive forces between like-charged particles, described via the Poisson-Boltzmann theory, entirely neglecting the fluctuation-induced forces, i.e. the interactions due to correlations in charge fluctuations and the effects of induced polarization charges at the dielectric discontinuities (i.e. the image charges).[55-61] As the origin of these fluctuation-induced forces can be described similar to the origin of the Casimir effect, they are often called as the "Casimir-like" forces. [58-59] Under certain conditions, as they can prevail in complex nanosystems, these Casimir-like forces attain a sufficiently large magnitude, such that the solution of the Poisson-Boltzmann equation remains no longer valid. For instance, intensive charge-fluctuations can turn the repulsive interaction between two like-charged particles into an attractive one[56-61]. Thus, to incorporate forces induced by the fluctuations of the counterions, the surface layer-charges, and the coupling of both, the traditional DLVO theory, as per the work of Lukatsky and Safran, was modified and termed as the "refined DLVO theory". [64] The related simulations can be applied for particle separations significantly smaller than the Goy-Chapman length, which is approx. 10 µm for our aqueous systems.

According to several estimations, measurements and force field calculations (for details, refer the supporting information), we concluded that the interactions among the semiconductor crystal and the underlying substrate surface can be best described by the DLVO model for the plate-like interactions[53]. The performed refined DLVO simulations, based on this model, consider solely $H_3O^+$ ions, the related concentrations being derived from the corresponding pH values. Now, why is the focus set on $H_3O^+$ ions? Answering this, it can be said that one of the major parameters affecting the DLVO simulation is the concentration of counterions, forming the diffuse part of the electrical double layer. Hence, considering the positively charged counterions being present in the different aqueous systems, as explored in this study, results revealed the consistent presence of $H_3O^+$, whereas the phosphate salts and the sodium chloride delivering $Na^+$, and KOH delivering $K^+$, respectively. These ions differing significantly in their spatial extensions, an OSWD can however only take place if a semiconductor crystal is located well within the operating distance of the π-π interactions that drive the deposition of the molecules during the OSWD process, i.e. within approx. 3.8 Å.[53]

Considering the dimensions of the phosphate salts (roughly calculated by adding up the Van-der-Waals diameters), the spatial extent of the smallest ion in this respect, disodium phosphate, was estimated to be 11.88 Å.[54] Hence, the two opposing double layers, consisting at least of two layers of phosphate ions and two layers of $Na^+$ - ions (4.54 Å), would have a smallest layer thickness of 32.84 Å, thereby making the probability of wide-stretching phosphate salts directing the OSWD very unlikely. Further, it was found that the structuring of water molecules around hydrophobic objects (such as γQAC particles and the HOPG substrate,[35-38]) prevents the adsorption of salt ions, as stated above. Additionally, the findings of molecular dynamics simulations show that small hard (non-polarizable) ions, such as alkali cations, are repelled from the surface of a hydrophobic object within a range of approx. 4 Å, in accordance with the standard Onsager–Samaras theory.[50-52] Nevertheless, simulations considering the actual salt concentration of the investigated aqueous samples, by applying both the traditional and the refined DLVO theory, were performed in addition; the results derived using the actual salt concentrations, however, depicting no correlation to the corresponding surface coverage. A detailed discussion of these simulations can be found in the supporting information.

The results of the appropriate, refined DLVO simulations considering solely $H_3O^+$ ions revealed that though for all the samples the curve progressions are similar in nature, the repulsive energy barrier level (i.e. the interaction energy maximum) of the sample modified with disodium phosphate is significantly higher than for all the other aqueous samples (Figure 9, (a)). In-depth analysis in this regard revealed that the repulsive energy barrier decreases exponentially with increasing pH (with all other conditions being kept identical), since the refined DLVO theory corresponds in mathematical terms to a logarithm function (a diagram depicting this finding can be found in the supporting information). As a result, a potential correlation between the interaction energy maxima (Figure 9 (b)) and the achieved surface coverage can not be evaluated by determining the coefficient of determination $R^2$ directly using linear regression methods. Consequently, a logarithm function was fit to the data as regression function and the regression function was transformed in a quasilinear function using logarithmic principles. The coefficient of determination thus obtained yields $R^2 = 47$ %, indicating a correlation between interaction energy barrier and surface coverage.

Results further revealed that modifying the catalyzing aqueous dispersion via ions does not only affect the zeta potential of the dispersed semiconductor particles, but also modifies the interaction forces between the semiconductor particle and the substrate.

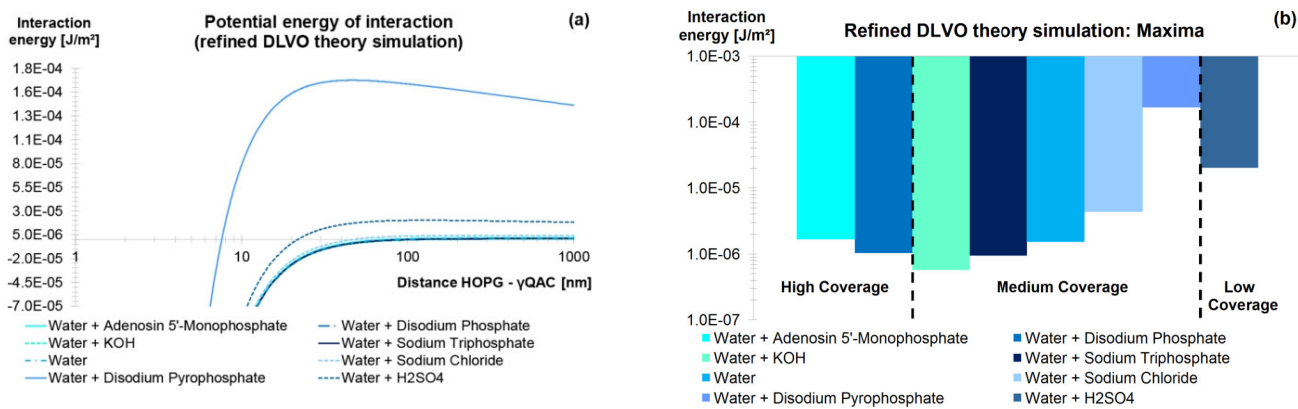

**Figure 9.** Refined DLVO simulations of the investigated aqueous samples , including fluctuation-induced forces (calculations done using solely the concentration of the $H_3O^+$ ions), with results depicting: (a) the complete interaction energy functions, and  (b) comparison of the interaction energy maxima .

**Force field calculations.** The fundamental question that required understanding for the development of an OSWD model is how a γQAC crystal geometrically interacts with the substrate surface during the OSWD process. The findings of previous experiments showed that the migration of the adsorbed QAC molecules (which is related to the surface diffusion) is limited up to temperatures of 160 °C.[19] Further, the interactions that drive the deposition of the molecules during the OSWD process have a limited operating range, with a maximum of approx. 3.8 Å for the case of π-π interactions.[53] This indicates that a distinct supramolecular QAC array is built by the detachment of a single crystal face of a γQAC crystal in contact to the substrate surface. To test the feasibility of the assumption of such a detachment, force field calculations were performed. This was done to compare the binding energies of the QAC molecules within different crystal faces of a γQAC nanoparticle with their interaction to a (0001) graphite surface, in order to investigate if γQAC crystal faces exist that contribute to the OSWD process. Such calculations allow for the simulation of relatively large chemical systems, such as complex organic crystals with sizes of several unit cells (Figure 10).

Initial calculations were performed using the Dreiding force field. Results revealed that a γQAC crystal comprises at least one crystal face ((010)) that would release molecules upon contact to the graphite substrate, owing to the comparative lower binding energy of the former (Figure 11). To estimate the accuracy of the obtained energy values, γQAC crystals were modeled with twice the number of γQAC unit cells in each of the directions (refer the supporting information). The results indicated an error of ± 1.5 kcal/mol, with all the deviations following a physical plausible trend: a slight decrease in the energy values with an increasing crystal size. For further verification of the results, two additional force fields for comparative studies were used: Universal Force Field (UFF) and Consistent Valence Force Field (CVFF). Comparison of these results with the Dreiding values revealed that the energy relation between the (010) crystal face of the γQAC and the graphite substrate could be reproduced with all the used force fields (Figure 11). Moreover, the energy values obtained via the Dreiding and the UFF force field have been found to be in the same size range as the published lattice energies of the γQAC.[62] This is in contrast to the CVFF force field calculations, which are well known to overestimate the cohesive energies of  the aromatic molecules (such as QAC) by 80% in comparison to the experimental values.[63]

Summing it up, the results of the force field calculations support the assumption that a distinct supramolecular QAC array is built by the detachment of a single crystal face of a γQAC crystal in contact with the graphite surface. However, the sum of our experimental and theoretical results requires participation of additional forces to the OSWD process, which plays an important role for proposing a suitable model of the OSWD. This is done next.

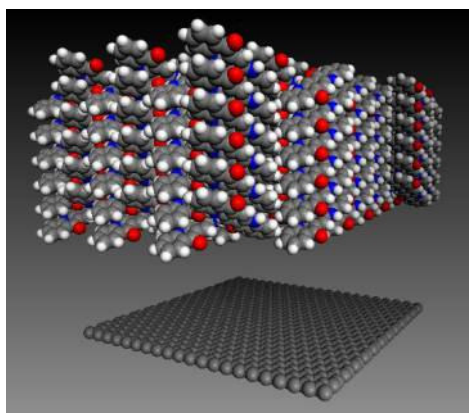

**Figure 10.** Force field calculated simulation of a γQAC crystal facing a HOPG substrate.

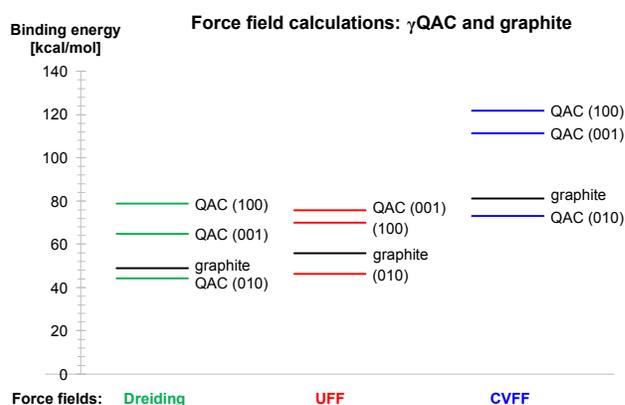

**Figure 11.** Results of the force field calculations. The image displays the calculated binding energies of a QAC molecule within specific crystal faces of a γQAC crystal (coloured lines), and of a single QAC molecule adsorbed on a (0001) graphite layer (black lines), derived from three different force fields, viz. the Dreiding, the UFF, and the CVFF.

## THE OSWD MODEL: GROWTH OF 2D ADSORBATES DIRECTLY FROM 3D PARTICLES

Based on the above results, the following model (Figure 12) is proposed: the dispersing agent charges the surfaces of both the dispersed organic crystals and the wetted substrate, thereby generating zeta potentials. The generated potential depends on the dispersing agent in use and further on the pH, in the case of aqueous dispersions.

**Crystal – surface distance ≤ 100 nm.** A dispersed organic crystal, approaching the substrate surface by Brownnian motion, encounters a repulsive barrier at a distance of about 100 nm. This barrier is generated by the double layer forces originating from the charge distribution around the surfaces. However, clearly negative charged crystals are able to overcome the repulsive energy barrier, since the level of the barrier decreases with an increasing negative zeta potential, according to our refined DLVO calculations. The interplay of the Poisson-Boltzmann related, Van-der-Waals, and fluctuation-induced (Casimir-like, respectively) forces leads to attractive interactions between the dispersed particles and the substrate surface. Further, such an attraction increases with a decrease in distance (between the organic crystal and the substrate surface), as revealed by the refined DLVO approach.

**Crystal – surface distance ≤ 10 nm.** At a distance of about 10 nm and below, the proposed model incorporates an additional force, arising from the hydrophobicity of both the substrate and the dispersed organic semiconductor crystals.[35-38] This additional force can be traced back to a phenomenon that is hardly accessed experimentally or by simulations, owing to its repercussions depending on numerous parameters: the dewetting in a hydrophobic confinement.[67-68] Such a confinement arises when two hydrophobic surfaces approach each other. At small separations of few nanometers, water or another polar organic liquid between the two hydrophobic surfaces becomes metastable (with respect to its vapor), and hence a dewetting transition is triggered, resulting in attractive interactions.[69-71] For the case of fully nonpolar dispersing agents, the expulsion of adsorbed liquid phase molecules was reported due to the collision among the dispersed particles (say, owing to the Brownian motion),[67,72] there being however a lack of appropriate publications analyzing this effect in detail. Nevertheless, from the analogous results, as presented, it can be strongly said that an associated dewetting effect prevails for the non-polar dispersing agents. Further, our STM results and the limited operating distance of the molecular orbital interactions exclude the presence of an additional, dispersing agent's adsorbate layer, between the semiconductor monolayer and the substrate surface. Thus, it can be proposed that during the approach of the semiconductor crystal towards the substrate, the liquid phase molecules and the weakly bonded ionic species within the diffuse layer are expelled by the hydrophobic dewetting effects. This is followed by the expulsion of the solvated ions constituting the outer Helmholtz planes of the involved surfaces (as per the Gouy-Chapman-Stern-Graham model of the electrical double layer).[44,66]

**Crystal – surface distance ≤ 0.5 nm.** The remaining non-solvated ionic adsorbates, forming the inner Helmholtz planes, induce a steric barrier at a separation of approx. 5 Å between the dispersed organic particles and the substrate. The results of the refined DLVO simulations indicated that when using aqueous dispersions, the inner Helmholtz planes contain a higher share of low-dimensional ionic species, as compared to the composition of the diffuse layer. This phenomenon is referred to the specific adsorption capacities of hydrophobic surfaces, as they generate a layer of ice-like structured water molecules that selectively adsorb hydroxyl ions, whereas other ions are repelled.[50-52] However, the sum of all the attractive forces, including the hydrophobic dewetting interaction, reaches such a level at this separation that the remaining ionic adsorbates of the inner Helmholtz planes are also expelled, thereby eliminating the steric barrier. Such a model was further found in accordance with the findings of corresponding experiments.[67,72] Hence, it can be proposed that a strong physical contact between the organic crystal and the substrate surface is established during the expulsion of the inner Helmholtz planes. Further, the simultaneous expulsion of the ionic adsorbates and the strong, direct physical contact between the semiconductor crystals and the substrate surface triggers the OSWD, if the facing semiconductor crystal plane exhibits a lower binding energy as compared to the substrate surface. Consequently, the semiconductor molecules detach from the semiconductor crystal and attach to the substrate surface.

An essential part of the molecule detachment can be attributed to the properties and conditions of the involved electrical double layers, provided by the inner Helmholtz layers. We propose in this regard that the double layer forces, including Poisson-Boltzmann related and fluctuation induced forces, distinctly strengthen the attraction towards the substrate surface. Hence, the gradient of the surface free energy is modified and thereby the detachment of the semiconductor molecules is catalyzed. In this respect, the zeta potential z33 was introduced as a comparative value, indicating that more negative is the z33 value, higher is the ultimately generated substrate surface coverage. Further, it can be said that the semiconductor molecules finally attach themselves to the substrate surface via non-covalent bonding, and for the case of a graphitic substrate, mainly via the π-π stacking. As a result, self-assembly processes are initiated, driven by intermolecular forces and diffusion processes, and directed via epitaxy, [65] leading to the assembly of supramolecular surface adsorbates. However, in accordance with the STM experimental findings, it can be said that the transfer and the assembly of the QAC arrays is completed within seconds or less.[17,20]

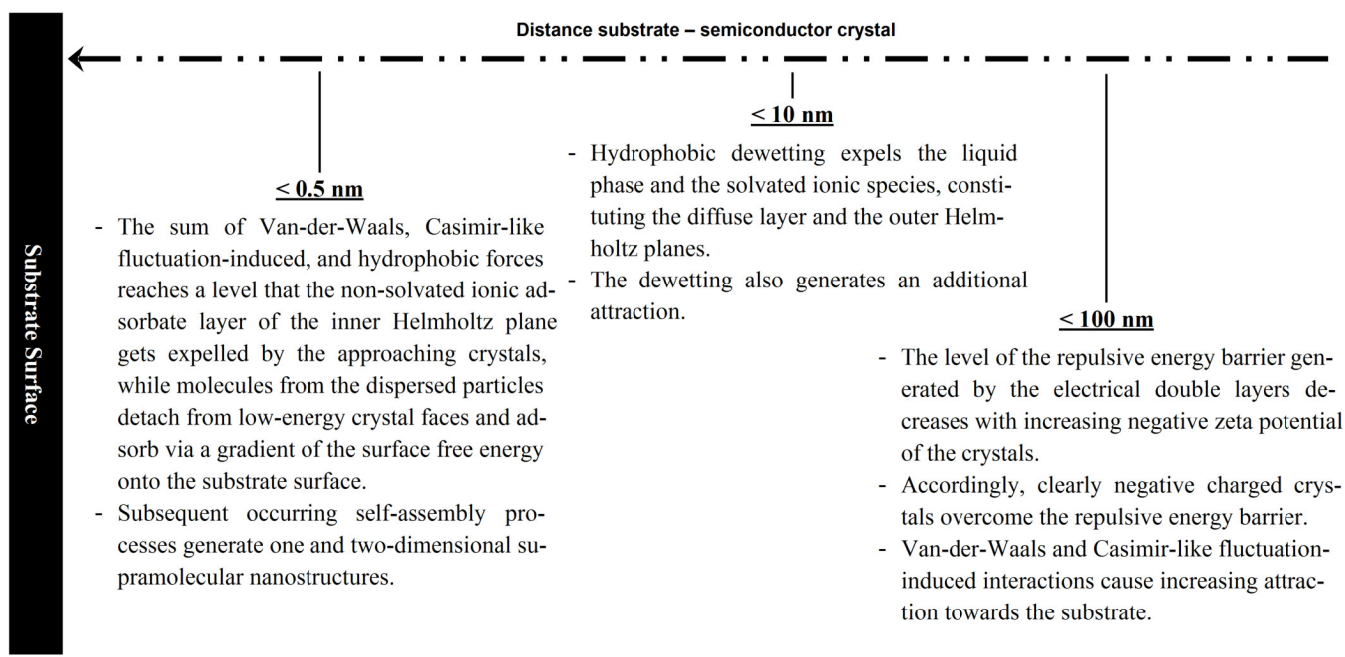

**Figure 12.** Summary of the major processes a semiconductor crystal goes through, as it approaches the substrate surface.

## CONCLUSIONS

Our OSWD model is based on the complex interaction of several basic forces, such as the Casimir-like fluctuation-induced forces, the dewetting-induced hydrophobic collapse, and the electric double layer forces. The model identifies the most important parameters affecting the OSWD process, most notable of them being the zeta potential. Changing the dispersing agent alters the zeta potential, and the latter hence allows the selection and modification of the dispersing agents accordingly, as to enable different processing technologies. This became especially workable for the aqueous dispersing systems, where adjusting the final surface coverage became straightforward, by the modification of the zeta potential using phosphate salts, acid or base. All the discovered possibilities of control are not locally restricted and do not require a subsequent rework of the fabricated sample. The identified characteristics of the OSWD point towards the fabrication of low-cost, but large-scale products, as the printed and potentially flexible carbon based electronics[23] or highly efficient systems to capture carbon dioxide.[4] Further, owing to the identification of water as a suitable dispersing agent, all kinds of non-toxic applications are thus imaginable.

## MATERIALS AND METHODS

**Refined DLVO simulation, including fluctuation-induced forces.** To include all kinds of fluctuation-induces forces, we adjusted the traditional DLVO simulation as per the work of Lukatsky and Safran.[64] This enhanced term for the calculation of the potential energy of interaction $V_{eT}$ consists of three sub-terms: one describing the attractive Van-der-Waals interaction, a second describing the unscreened Poisson-Boltzmann repulsion, and a third treating the effect of fluctuation-induced forces, expressed as:

$$V_{eT}(h) = -\frac{H_{123}}{12\pi} * \frac{1}{h^2} - \frac{kT}{\pi\lambda l_B}\ln(h) + \frac{7kT}{4\pi\lambda^2}\ln(h) \qquad (1)$$

Where $h$ is the separation between the HOPG and the γQAC surface; $H_{123}$ is the Hamaker constant for the system HOPG (index 1), dispersing agent (index 2), and γQAC (index 3); $k$ is the Boltzmann's constant; $T$ is the temperature; $\lambda$ is the Gouy-Chapman length; and $l_B$ is the Bjerrum length.

Further detailed information describing the single calculations and parameters necessary to perform the refined DLVO simulations can be found in the supporting information.

## ASSOCIATED CONTENT

**Supporting Information**

Experimental details concerning the STM, the zeta potential, the particle size, and the pH measurements. STM pictures related to all the tested dispersing agents, of how to distinguish monolayer from bilayer adsorbate structures, and related to STM measurement artifacts. Detailed correlation analysis related to the physical properties of the examined dispersing agents and scatter plots relating the surface coverage to the particle size, the zeta potential, the pH, and the interaction energy maximum of the investigated systems. Experimental details and results concerning scanning electron microscopy measurements. Further information about the charging mechanism in aqueous systems, leaky dielectrics and for fully non-polar media. Detailed description on how the surface coverage rate was determined. The specifications of the performed force field calculations. Detailed information on how the $\gamma$QAC crystal size was calculated via a back calculation and how a suitable DLVO model was chosen. A detailed description of the single calculations and the parameters necessary to perform the traditional, as well as the refined DLVO simulations. A detailed discussion of the performed traditional DLVO simulations and further results related to the refined DLVO simulation, including a comparative study exploring the influence of the pH on the DLVO calculations. This material is available free of charge via the Internet at http://pubs.acs.org.

## AUTHOR INFORMATION

**Corresponding Author**

*E-mail: Trixler@lrz.uni-muenchen.de

**Author Contributions**

The manuscript was written through contributions of all authors.

**Funding Sources**

The Bayerisches Staatsministerium für Umwelt und Verbraucherschutz and the mentoring program of the Nanosystems Initiative Munich (NIM) is sincerely acknowledged for their funding.

**Notes**

The authors declare no competing financial interest.

## ACKNOWLEDGMENT

The authors would like to thank Martina Zeiler from the Papiertechnische Stiftung in München for her kind support regarding the zeta potential measurements. Futhermore, we'd like to thank Veronika Schömig from the Technische Universität München, Fachgebiet für Selektive Trenntechnik, for her quick and smooth support, enabling us to perform a series of particle size measurements. Big thanks as well go to Manfred Seidel from the Deutsche Museum, München, for all his aid in the field of chemistry and much more. Also, we'd like to acknowledge Klaus Macknapp from the Deutsche Museum, München, for his help and support regarding the SEM analysis. Further, we would like to thank Neeti Phatak for her support with the proofreading and editing of the document. We additionally would like to thank our colleague Andrea Greiner, who came up with the idea of displaying the determined surface coverage via a box plot diagram. Last, but certainly not the least, we'd like to thank Wolfgang W. Schmahl and Guntram Jordan for their excellent idea of performing DLVO theory related simulations.

## ABBREVIATIONS

OSWD, Organic Solid-Solid Wetting Deposition; 1D, 2D, 3D, one-, two-, three-dimensional; QAC, quinacridone; HOPG, highly oriented pyrolytic graphite; STM, Scanning Tunneling Microscopy; z50, the median value of a distribution; z33, the point where 33 % of a distribution is more negative and 66 % is more positive; SC, surface coverage; R, the Pearson correlation coefficient; Ř, the coefficient of determination.

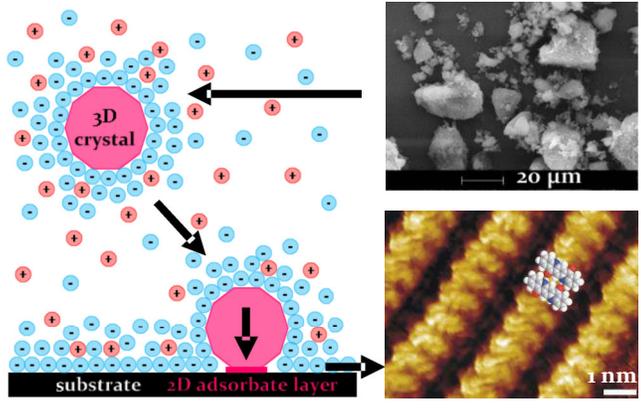

# Supporting Information:

# Revealing the Physicochemical Basis of Organic Solid-Solid Wetting Deposition: Casimir-like Forces, Hydrophobic Collapse, and the Role of the Zeta Potential

Alexander Eberle,[†,‡] Thomas Markert,[§] and Frank Trixler*[,†,‖]

[†]Department of Earth and Environmental Sciences and Center for NanoScience (CeNS), Ludwig-Maximilians-Universität München, Theresienstraße 41, 80333 München, Germany

[‡]Helmholtz Institute Erlangen-Nürnberg for Renewable Energy, Egerlandstr. 3, 91058 Erlangen, Germany

[§]Institute of Theoretical Chemistry, Ulm University, Albert-Einstein-Allee 11, 89081 Ulm, Germany

[‖]TUM School of Education, Technical University of Munich and Deutsches Museum, Museumsinsel 1, 80538 München, Germany

## Table of Contents

## 1. STM sample preparation and scan setting

To prepare a standard scanning tunneling microscope (STM) sample, using a dispersion with 2 wt. % of pigment, 82 mg of powdered γQAC (5,12-Dihydro-quino[2,3-b]acridine-7,14-dione, CAS 1047-16-1, purchased as Hostaperm Red E5B02 from Clariant) was dispersed in 4 ml of the desired dispersing agent. In this regard, the dispersing agents used were: propane-1,2-diol (CAS 57-55-6, from Sigma Aldrich, item no. 82280), 3-methyl-1-butanol (CAS 123-51-3, from Sigma Aldrich, item no. 59090), acetone (CAS 67-64-1, from Sigma Aldrich, item no. 34850), anisole (CAS 100-66-3, from Sigma Aldrich, item no. 10520), dimethyl sulfoxide (CAS 67-68-5, from Sigma Aldrich, item no. 472301), ethyl acetate (CAS 141-78-6, from Sigma Aldrich, item no. 45767), ethylbenzene (CAS 100-41-4, from Sigma Aldrich, item no. 03080), 2-Propanol (CAS 67-63-0, from Sigma Aldrich, item no. 34965), propylene carbonate (CAS 108-32-7, from Sigma Aldrich, item no. 310328), dodecane (CAS 112-40-3, purchased from Sigma Aldrich, item no. D221104), and purified water. For some cases where aqueous samples were used, additionally 0.5 wt. % salt (corresponding to 21 mg for a sample having a volume of 4 ml) was added to the sample mix. Tests were conducted using the salts sodium triphosphate (pentabasic, CAS 7758-29-4, from Sigma Aldrich, item no. 72061), disodium phosphate (dihydrate, CAS 10028-24-7, purchased as di-Natriumhydrogenphosphat-Dihydrat, from Merck Millipore, item no. 1065801000), disodium pyrophosphate (purchased as sodium pyrophosphate dibasic, CAS 7758-16-9, from Sigma Aldrich, item no. 71501), sodium chloride (CAS 7647-14-5, unknown supplier, but purity determined via XPS analysis: ≥ 99,99 %), and adenosine 5'-monophosphate disodium (salt, CAS 4578-31-8, from Sigma Aldrich, item no. 01930). At times, when mentioned, 1 µl of a 2 M potassium hydroxide (KOH) solution (CAS 1310-58-3, pellets purchased from Carl Roth, item no. P747.1), or 2 µl of a 1 M sulfuric acid ($H_2SO_4$) solution (CAS 7664-93-9, from Walter CMP, item no. 016-020-00-8), was alternatively added.

To begin with, a few drops of the dispersion were dispensed on a highly ordered pyrolytic graphite (HOPG, purchased from NT-MDT, item no. GRBS/1.0). If the dispersing agent didn't vaporize after an exposure time of 10 – 20 minutes, as in the case of propane-1,2-diol, 3-methyl-1-butanol, anisole, dimethyl sulfoxide, ethylbenzene, propylene carbonate, and purified water, the HOPG was dried by a special hotplate, enabling an accurate temperature control and providing a smooth temperature increase (Stuart SD160, temperature accuracy ± 1.0 °C). Once the sample was dried, it was immediately taken off the hotplate. Note, that the heating up of pure γQAC powder to 160 °C without a dispersing agent was unable to induce an OSWD,[1-2] and all the dispersing agents vaporized below 160 °C. Further, creation of two-dimensional QAC arrays, without a dispersing agent, was only possible by heating pure γQAC powder to 240 °C.[1-2]

The ready-made STM samples being investigated within days; as per the previous tests, QAC arrays were revealed to not change their structure for a minimum of four weeks, provided they are not influenced via any external forces.[1] For the STM measurements, we used a home-built STM combined with a SPM 100 control system, supplied by RHK Technology. The scans settings were: bias = 1 V, tunnel current 300 = pA, and the line time = 50 ms. Further, the voltage pulses used to improve the scan quality were located in the range of 4.3 and 10 V. The STM measurements being performed under ambient conditions, a thin layer of dodecane was generally applied on top of the HOPG surface to increase the measurement quality.[6] Extensive tests revealed, that a mixture of dodecane and γQAC, without another dispersing agent, was unable to generate two-dimensional QAC structures and that the dodecane by itself could not form supramolecular assemblies at room temperatures.[7] Besides, to conduct a STM measurement under ambient conditions without dodecane was very challenging, mainly due to the occurrence of a thin contamination layer of adsorbed water on top of the HOPG surface,[3-5] which otherwise was found to be successfully removed using the hydrophobic dodecane.

**2. Further STM example pictures**

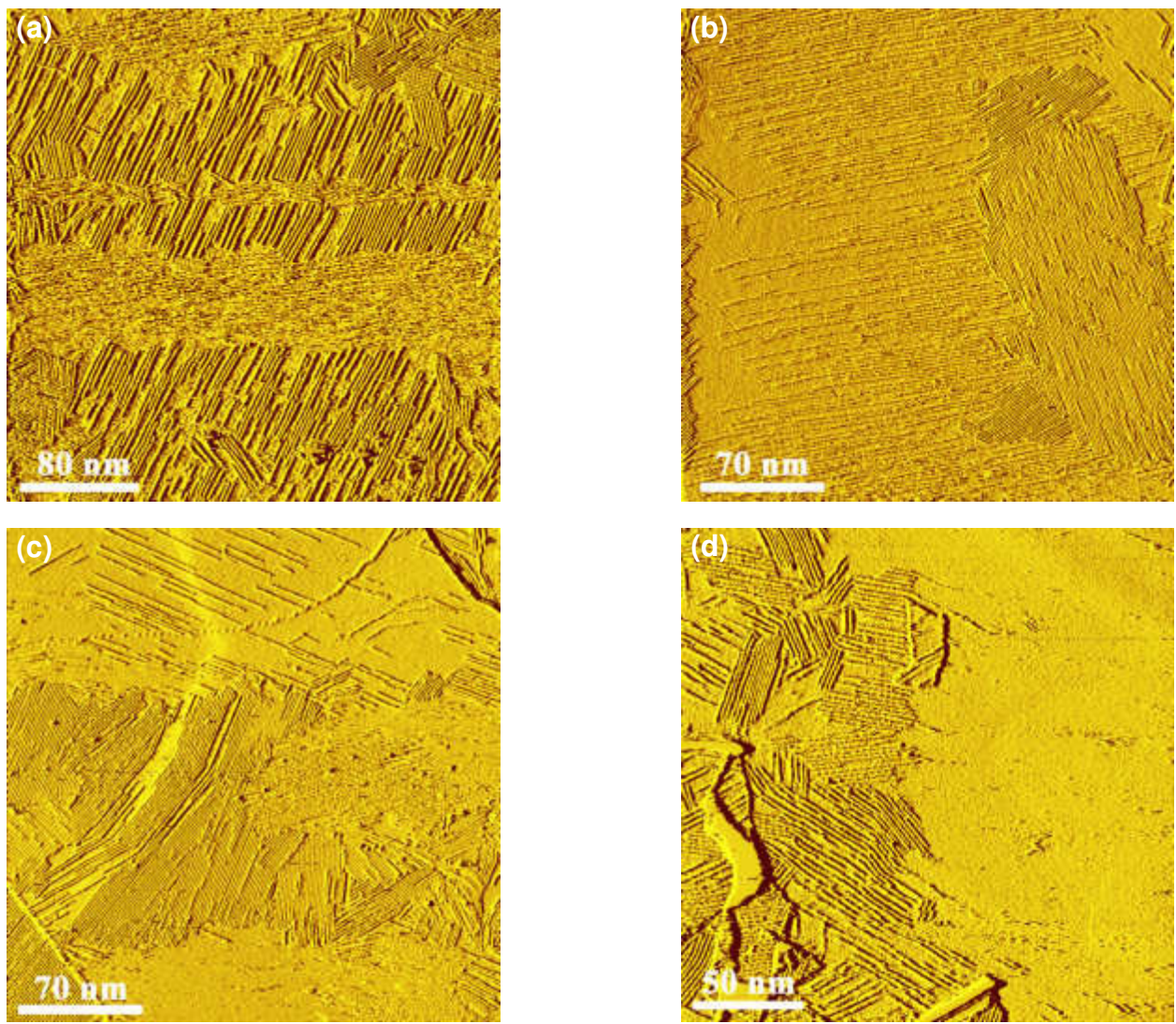

**Figure S1.** STM images showing exemplary the HOPG surface coverage (SC) by QAC adsorbates, depending on the dispersing agent in use. The surface coverage classified as the "High Coverage": (a) Propylene carbonate, SC = 75 ± 19 %; (b) Ethyl acetate, SC = 68 ± 15 %; (c) Anisole, SC = 61 ± 9 %; and (d) 2-Propanol, SC = 41 ± 14 %.

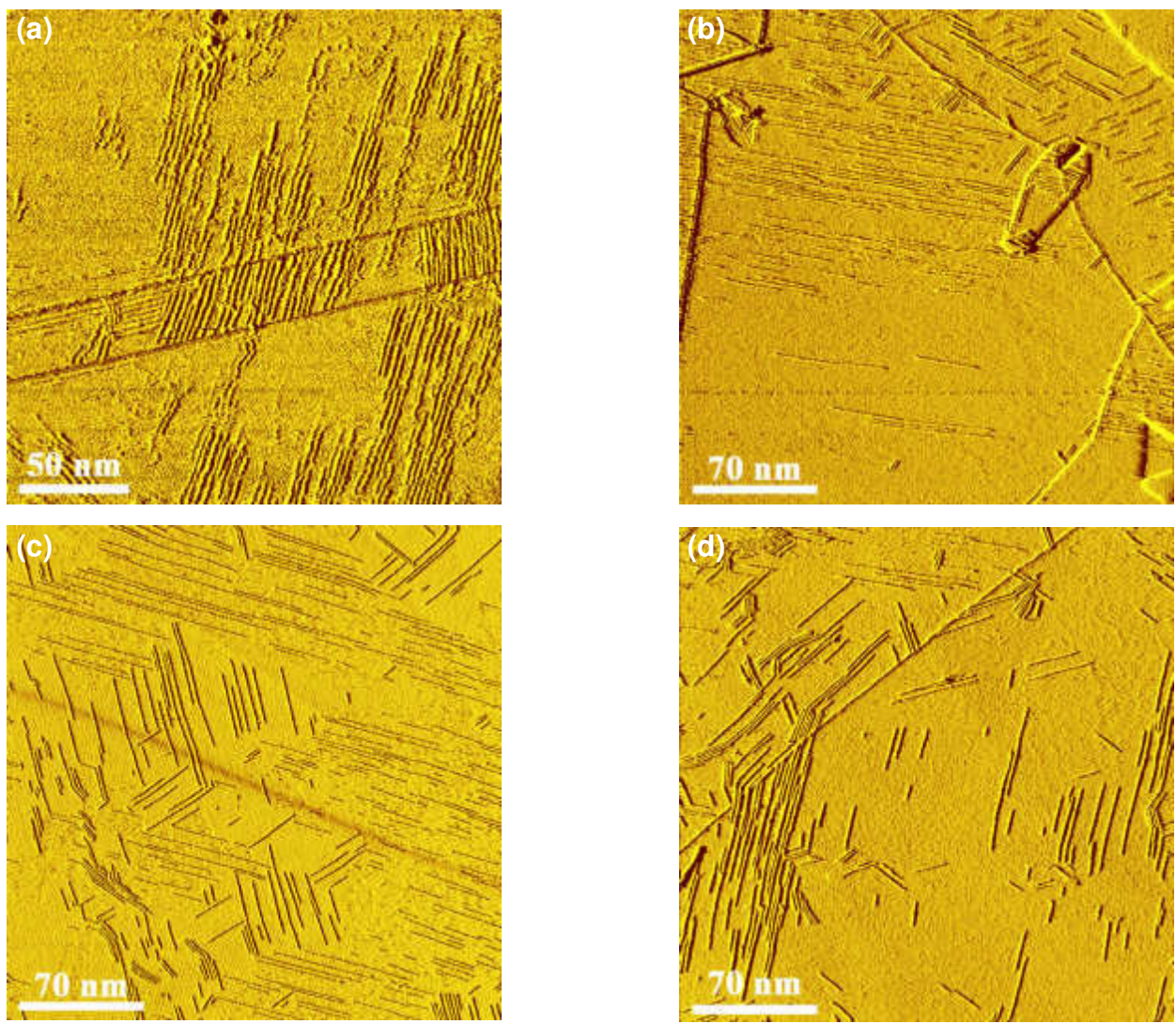

**Figure S2.** STM images showing exemplary the HOPG surface coverage (SC) by QAC adsorbates, depending on the dispersing agent in use. The surface coverage classified as the "Medium Coverage": (a) Acetone, SC = 38 ± 7 %; (b) Purified Water, SC = 24 ± 7 %; (c) Ethylbenzene, SC =22 ± 4 %; and (d) Propane-1,2-diol, SC = 21 ± 5 %.

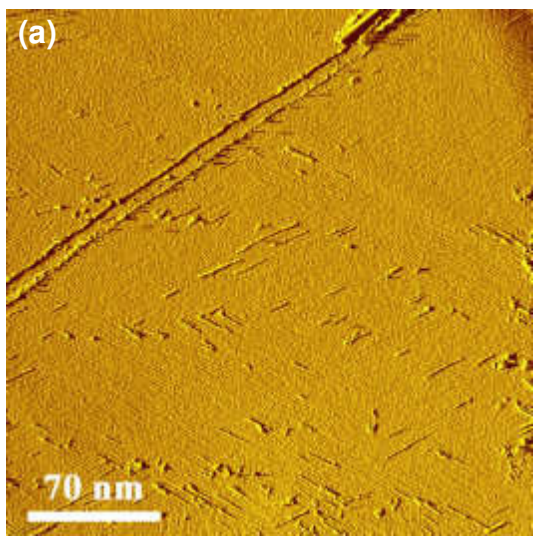

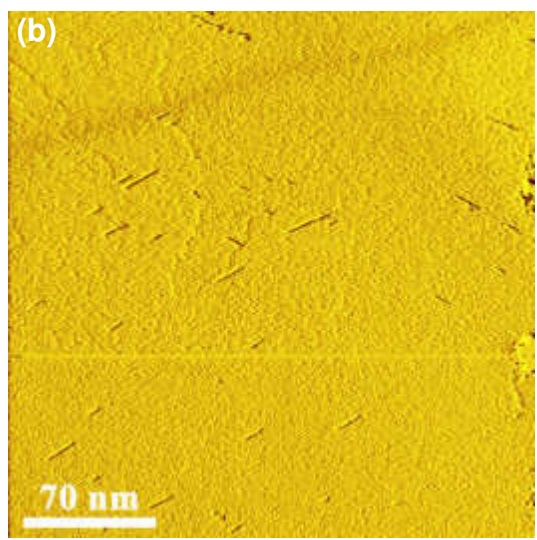

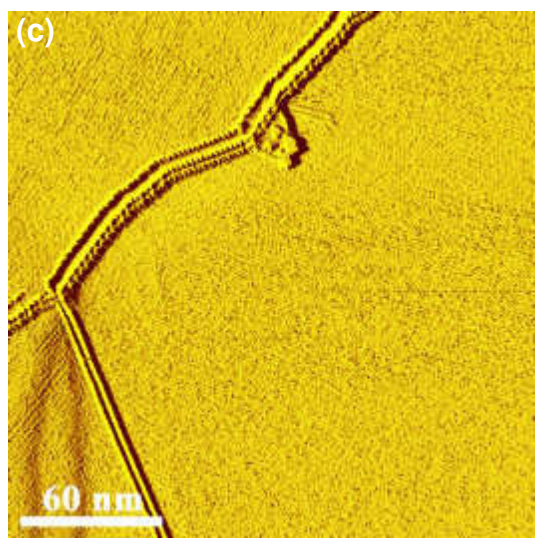

**Figure S3.** STM images showing exemplary the HOPG surface coverage (SC) by QAC adsorbates, depending on the dispersing agent in use. The surface coverage classified as the "Low Coverage": (a) Dimethyl sulfoxide, SC = 6 ± 4 %; (b) 3-Methyl-1-Butanol, SC = 0.7 ± 0.7 %; and (c) Dodecane, SC = 0.3 ± 0.3 %.

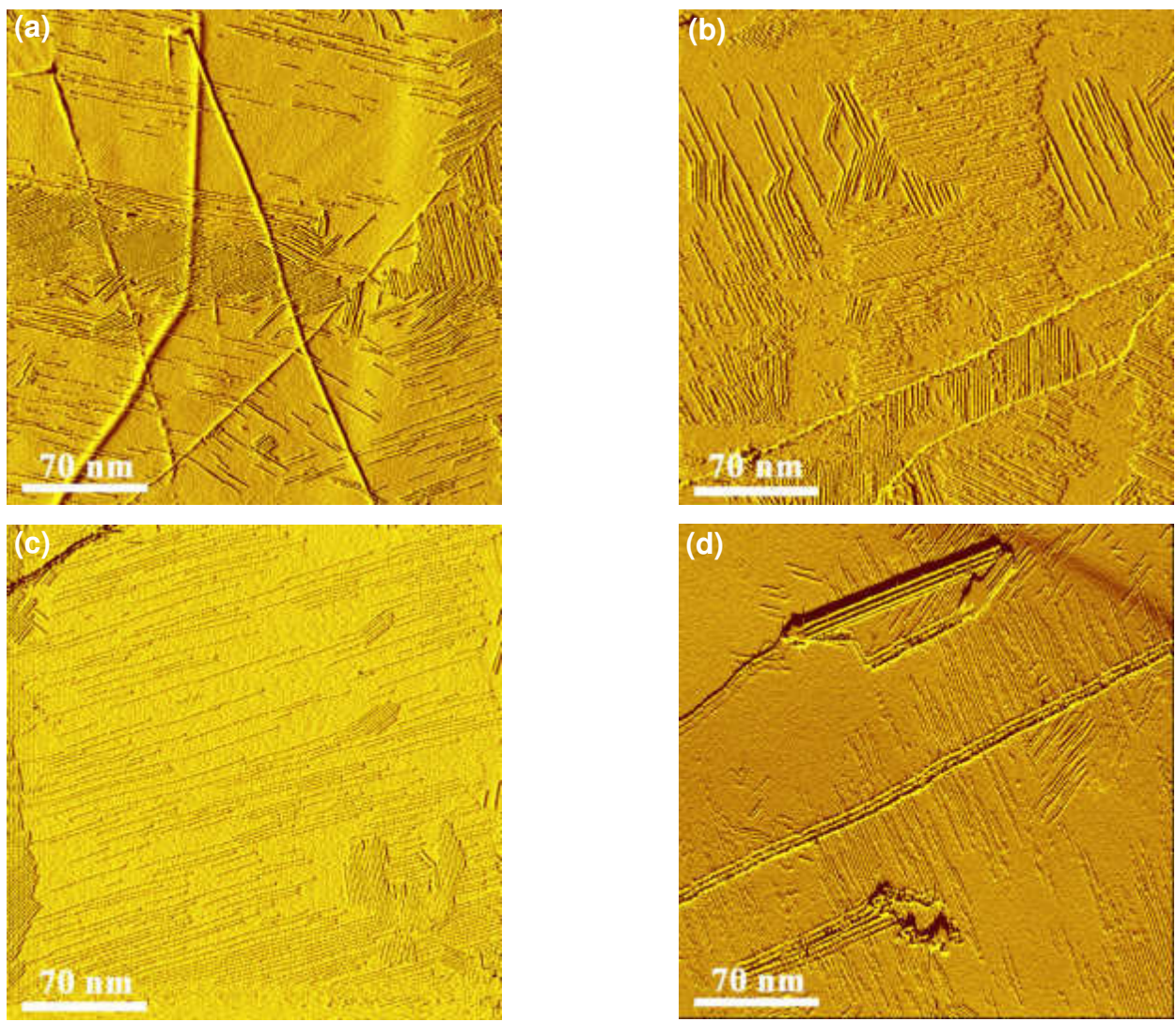

**Figure S4.** These STM example pictures show QAC surface adsorbates, generated atop a HOPG surface via aqueous γQAC dispersions that have been modified by adding different salts or the base KOH. Classified as the "High Coverage": (a) Adenosine 5'-monophosphate disodium, SC = 43 ± 11 %; and (b) Disodium phosphate, SC = 41 ± 13 % Classified as the "Medium Coverage": (c) KOH, SC = 38 ± 14 %; and (d) Sodium triphosphate, SC = 30 ± 19 %.

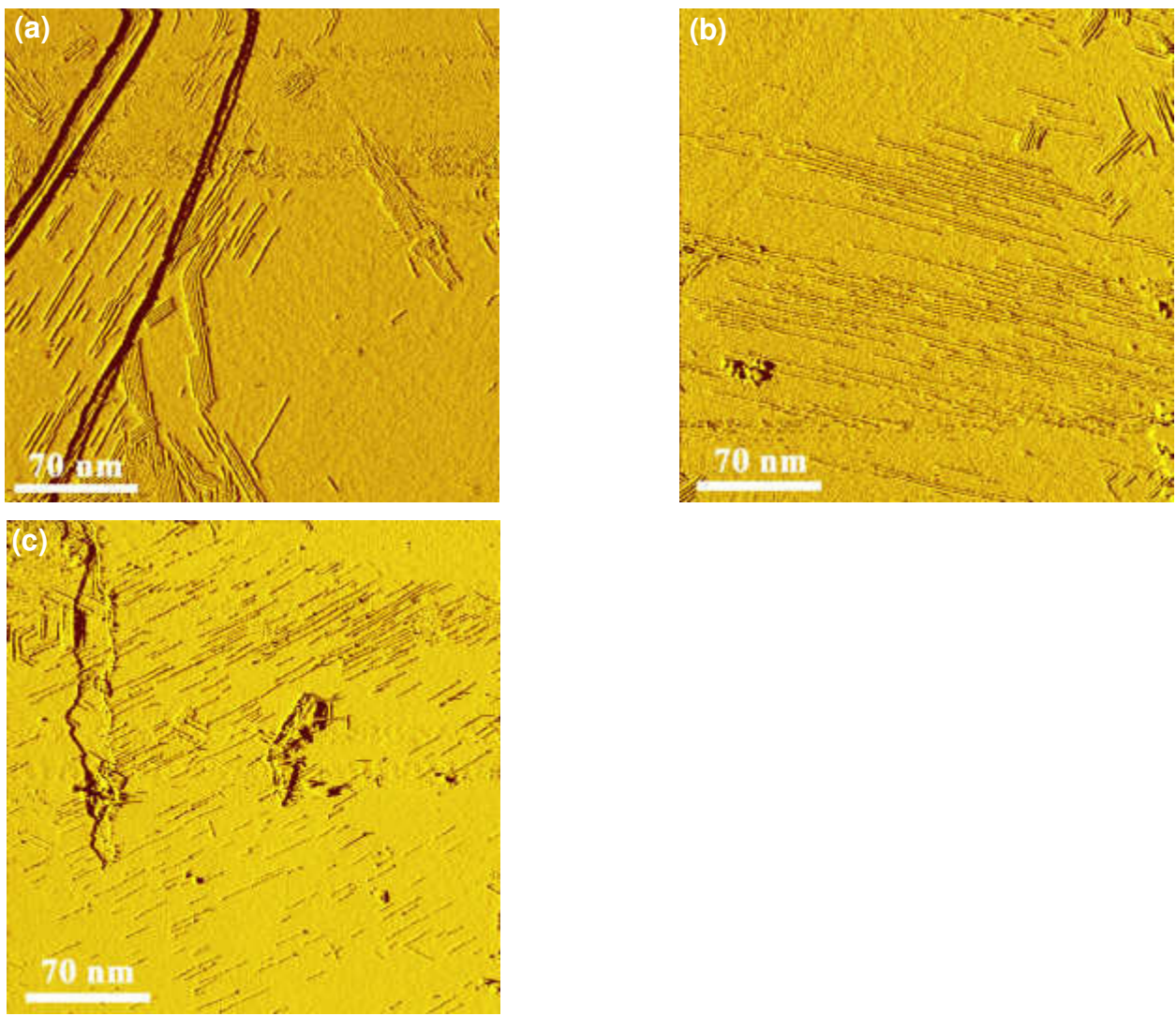

**Figure S5.** These STM example pictures show QAC surface adsorbates, generated atop a HOPG surface via aqueous γQAC dispersions that have been modified by adding different salts or the acid H$_2$SO$_4$. Classified as the "Medium Coverage": (a) Disodium pyrophosphate, SC = 17 ± 10 %. Classified as the "Low Coverage": (b) Sodium chloride, SC = 12 ± 13 %; and (c) H$_2$SO$_4$, SC = 11 ± 7 %.

### 3. Distinction between monolayer and bilayer adsorbate structures via STM

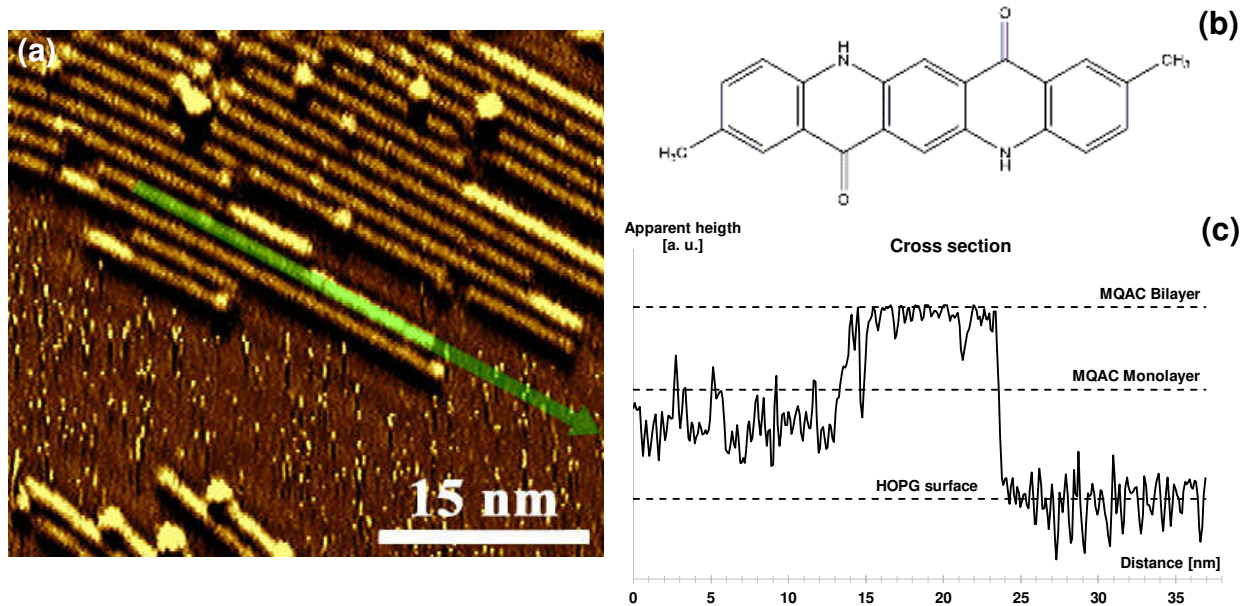

**Figure S6.** (a) STM picture of methylated quinacridone (MQAC) adsorbate structures. MQAC has almost the same chemical structure as QAC, but exhibits two additional methyl groups that support the formation of bilayer structures via steric interactions,[2] making thus the location and analysis of such bilayer structures much easier as compared to QAC bilayer structures. The linear features within the picture represent 1D MQAC chains that are built in the same way as supramolecular QAC chains are formed.[2] The significantly brighter sections atop of these linear features indicate plane-parallel bilayer structures, built as a supramolecular MQAC chain is formed directly on top of another one. (b) Chemical structure of MQAC. (c) Cross section of a plane-parallel MQAC bilayer, taken in the area marked with a green arrow in (a). Note that the apparent height is declared in arbitrary units, since the apparent height determined via STM depends on the STM scan settings, making accurate topographical measurements hardly possible.

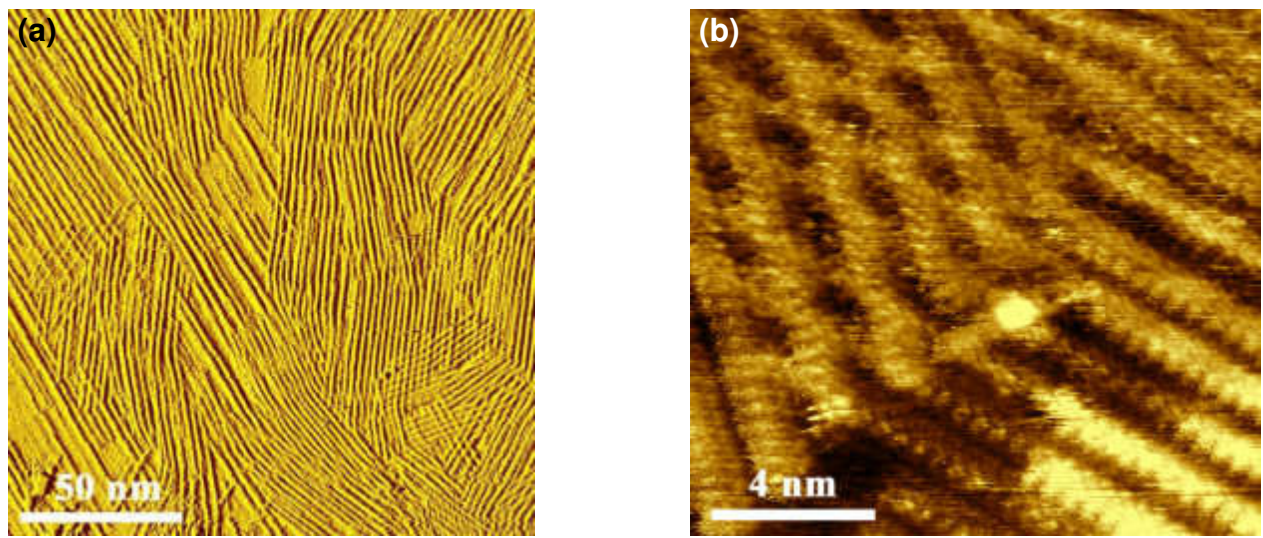

**Figure S7.** (a) STM picture of supramolecular QAC adsorbates atop a HOPG substrate. Two overlapping QAC arrays forming a bilayer exhibiting a crisscross structure can e.g. be seen in the bottom right corner. Note that it is hard to find QAC bilayer structures, making a detailed analysis challenging.[1] (b) Close-up view of QAC bilayer with a crisscross structure.

## 4. STM measurement artifacts

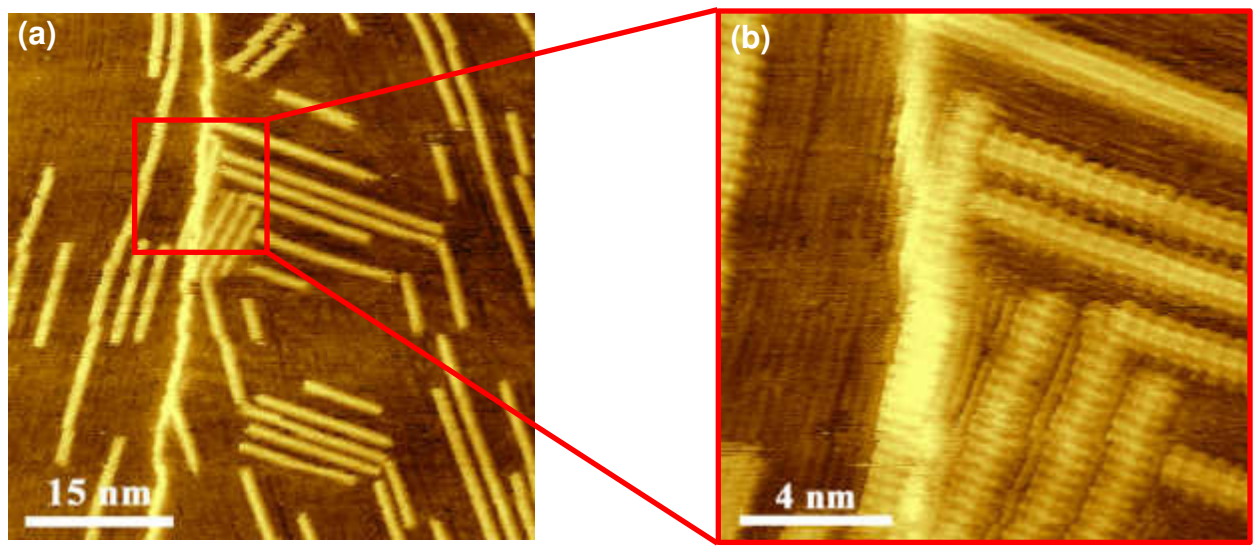

**Figure S8.** (a) STM picture of supramolecular QAC adsorbates on a HOPG. Additionally, chain-like structures can be seen next to the QAC structures, however, these structures are neither related to surface adsorbates nor to the substrate surface, but are related to measurement artefacts, them being the result of a poor quality of the STM tip and the ambient measurement conditions. Such measurement artefacts can be determined easily, since their size does not scale correctly: an increase in the magnification factor does not increase the size of these structures, as can be seen in (b). Further, related experiments have shown that these structures cannot be removed mechanically (by lowering the tip – substrate – distance), in contrast to actual adsorbates (like QAC molecules).

**5. Determining the surface coverage**

To determine the coverage of the HOPG surface by the QAC arrays within a single STM picture, the software Gwyddion (64bit), version 2.42 was used. For this, initially, the supramolecular QAC adsorbates (i.e. 1D chains and 2D arrays) via the tool "Mask Editor" were highlighted, followed by the export of single array dimensions by the tool "Grain distributions", finally accompanied by the Microsoft Excel 2013 calculations to determine the coverage ratio. Further, to investigate the average coverage of a STM sample, we analyzed per sample an area of about 0.7 µm², using a number of STM pictures with high scan resolution and without measurement artefacts, randomly selected from at least 5 clearly separated positions of the covered substrate; the average coverage rates including the double standard deviations being specified in the current publication. Calculations revealed high doubled standard deviations, indicating the dependence of the coverage by QAC arrays on different positions of a STM sample. Nevertheless, the average coverage rates of different samples treated with different dispersing agents were observed to differ significantly higher in value.

**6. Correlating physical properties of the dispersing agents to the surface coverage**

Referring to the article text, an analysis was performed to reveal a potential correlation between specific physical parameters of the dispersing agents in use and the achieved surface coverage. Going into detail, contribution of the viscosity to the OSWD process was expected, since the viscosity appeared to influence the mobility of the dispersed QAC crystals, and hence their ability to contact the HOPG surface (Figure S9). Further, a correlation between the dispersing agents' relative permittivity, indicating the dispersing agents' polarity, and the final surface coverage with QAC arrays was expected, since the polarity probably modifies the gradient of the surface free energy (Figure S10). Additional analysis was performed, since during the sample preparation, it was eye-catching to see that the dispersing agents spanned a huge range of different surface tensions and vapor pressures. Whereas some dispersions formed stable drops, others were seen spreading over the whole HOPG surface, and the γQAC powder was found partially forming an independent phase, especially when water was used as the dispersing agent. Further, whereas some dispersing agents were seen evaporating at room temperatures within minutes, others required a significant longer time to do so. For the related results, refer to Figure S11 and Figure S12.

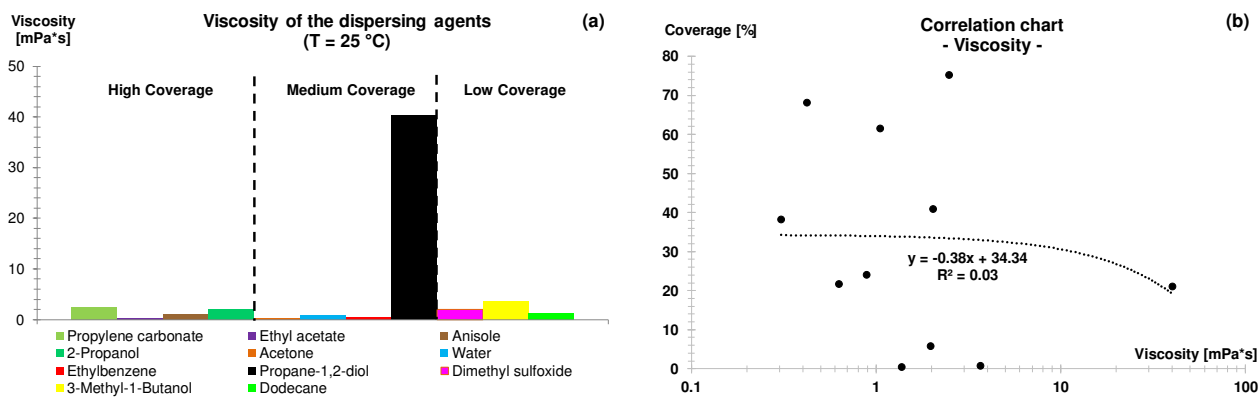

**Figure S9.** (a) The viscosity of the catalyzing dispersing agents in use,[8,9] classified according to the achieved surface coverage. (b) Scatter plot relating the median of the achieved surface coverage to the dispersing agent's viscosity. To avoid confusion, note that the nonlinear appearance of the regression line is related to the logarithmical scale of the x-axis.

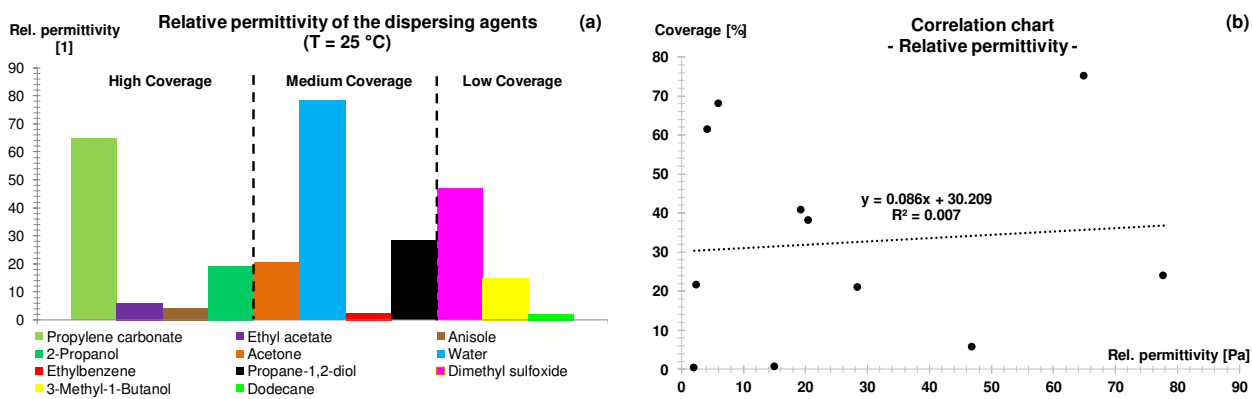

**Figure S10.** (a) The relative permittivity of the catalyzing dispersing agents in use,[9] classified according to the achieved surface coverage. (b) Scatter plot relating the median of the achieved surface coverage to the dispersing agent's relative permittivity.

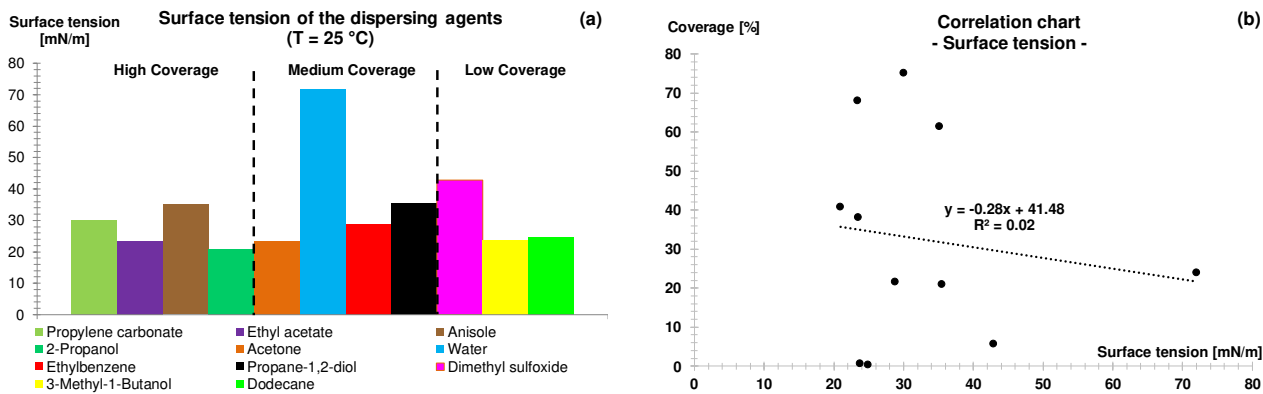

**Figure S11.** (a) The surface tension of the catalyzing dispersing agents in use,[9,10] classified according to the achieved surface coverage. (b) Scatter plot relating the median of the achieved surface coverage to the dispersing agent's surface tension.

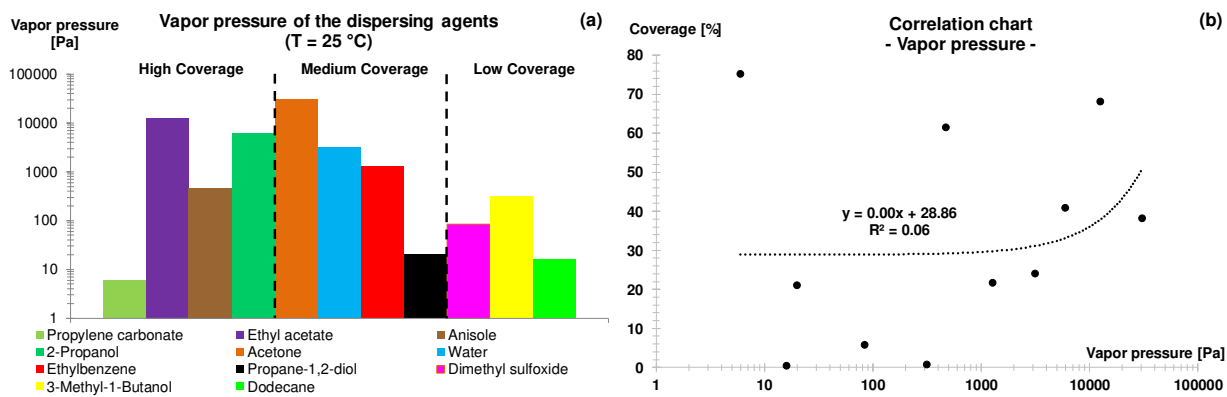

**Figure S12.** (a) The vapor pressure of the catalyzing dispersing agents in use,[9,11] classified according to the achieved surface coverage. (b) Scatter plot relating the median of the achieved surface coverage to the dispersing agent's vapor pressure. To avoid confusion, note that the nonlinear appearance of the regression line is related to the logarithmical scale of the x-axis.

## 7. Particle size measurements

For these measurements, we used a high-end dispersion analyzer, LUMiSizer 6112-95 from LUM. Further, for the analysis, 400 µl of sample, containing 0.5 wt. % pigment, and in some cases, additional of 0.125 wt. % salt, was used.

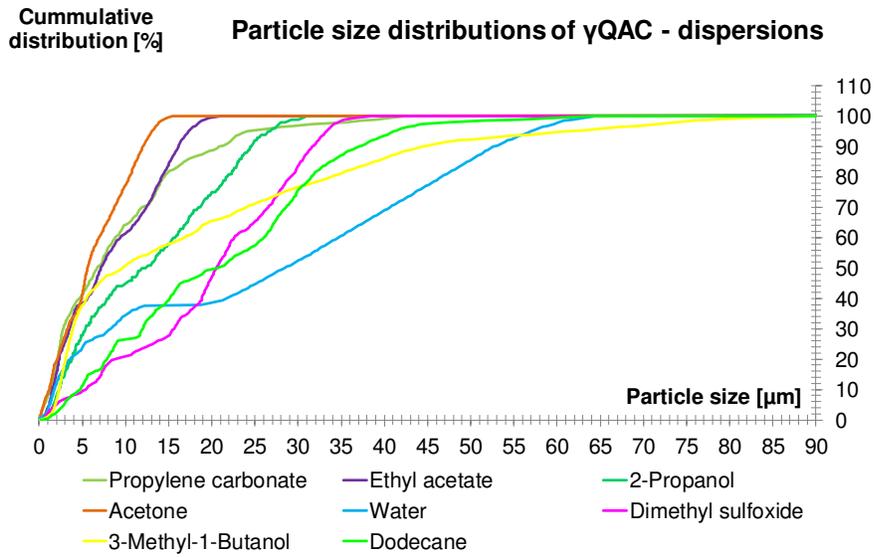

**Figure S13.** The cumulative distributions of the analyzed γQAC dispersions as a function of the particle size.

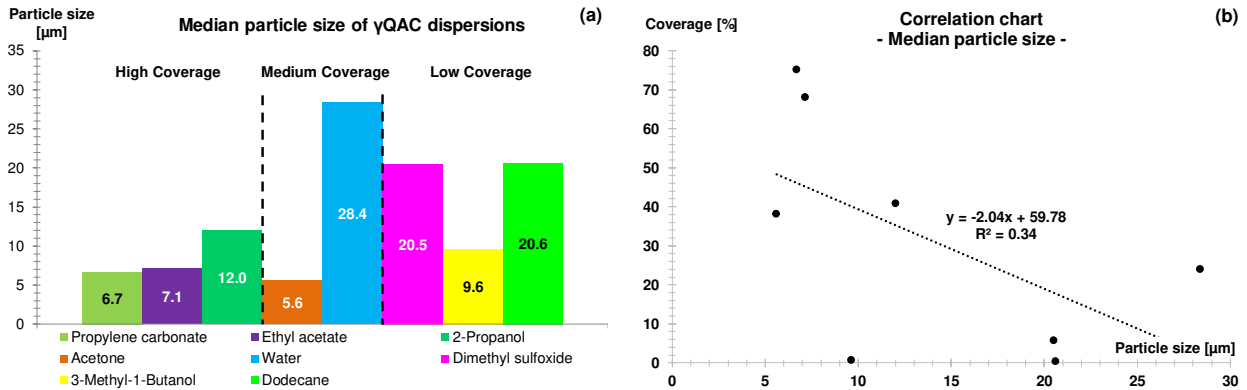

**Figure S14.** **(a)** Median particle size (equivalent diameter) of powdered γQAC, dispersed in different dispersing agents. **(b)** Scatter plot relating the median of the achieved surface coverage to the median of the particle size distribution.

**8. Zeta potential measurements**

All the zeta ($\zeta$) potential measurements were performed using a Zetasizer Nano ZS ZEN 3600 system from Malvern; 4 ml samples with 0.1 wt. % of $\gamma$QAC or graphite powder (from Alfa Aesar, item no. 14735) being used in this regard. Further, to determine the $\zeta$ potential distribution, three measuring cycles (each measuring cycle comprising of 30 single measurements) were performed per sample. The averaged results of a measuring cycle as a distribution of $\zeta$ potentials $z_{mi}$ were displayed by the Zetasizer Nano ZS ZEN 3600, where $z_{mi}$ represents the mean value of a $\zeta$ potential interval $i$, and the related intensities $I_i$ (i.e. the total counts per $\zeta$ potential interval). This kind of representation corresponds to the type of display used for particle size distributions.[12] Further, the relative quantity $\Delta I(z_{mi})$ is determined via the total intensity $I_{total}$, as:

$$\Delta I(z_{mi}) = \frac{I_i}{I_{total}} \tag{1}$$

The sum of these relative values $\Delta I(z_{mi})$, starting with the most negative $\zeta$ potential $z_{mmin}$ and going until $z_{mi}$, leads to the cumulative distribution $I(z_i)$, as:

$$I(z_i) = \sum_{z_{mmin}}^{z_{mi}} \Delta I(z_{mi}) \tag{2}$$

Worth noting here is that $z_i$ is the upper value of the $\zeta$ potential interval $i$, due to plotting of the cumulative distribution against the upper interval value.[12] Further, Figure S15 shows the averaged cumulative distributions of the analyzed $\gamma$QAC dispersions, derived by averaging 90 single $\zeta$ potential measurements per sample. Referring to the explanations in the article, both the z50 (the median value of the cumulative $\zeta$ potential distribution) and the z33 (the point where 33 % of the distribution is more negative and 66 % of the distribution is more positive) were derived from the corresponding cumulative distribution, employing linear interpolation, as:

$$zNN(I) = z_{i-1} + \frac{z_i - z_{i-1}}{I_i - I_{i-1}}(I - I_{i-1}) \tag{3}$$

Where $zNN$ is the desired $\zeta$ potential, $I$ the corresponding value of the cumulative distribution, the index $i$ being related to the upper value of corresponding interval, and the index $i-1$ being related to the lower value of the corresponding interval. Note: the double standard deviation was employed to estimate the quality of the derived value. Further, as the distribution-measurements failed for the dispersing agents dodecane and 3-methyl-1-butanol, a monomodal measurement mode that determines only the median value had to be used. Though these monomodal measurements are just roughly comparable to the other measurements, they however indicated that the $\zeta$ potential of the $\gamma$QAC crystals in dodecane and in 3-methyl-1-butanol is reverse to the $\zeta$ potentials in the other dispersing agents.

A comparison of the median values of the distributions (z50) indicated that the samples with a high z50 have a tendency of generating a high surface coverage, as can be seen in the Figure S16, however, the determined coefficient of determination $R^2 = 38$ % yielded no significant correlation to the surface coverage in this respect (in contrast to the z33 value, being discussed in the article text).

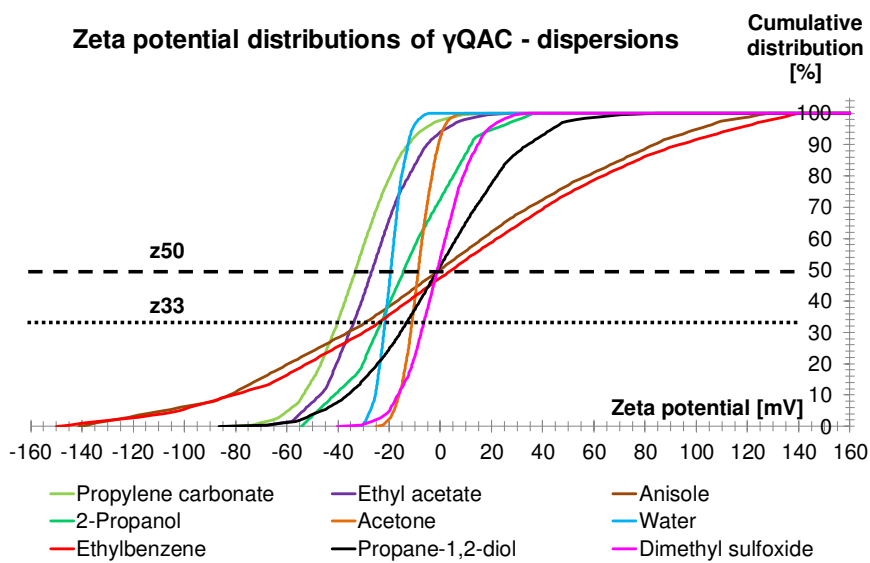

**Figure S15.** The cumulative distributions of the analyzed γQAC dispersions as a function of the ζ potential. The intersection of a sample's cumulative distribution and the dashed line yields the median value z50, while the intersection with the dotted line yields the z33.

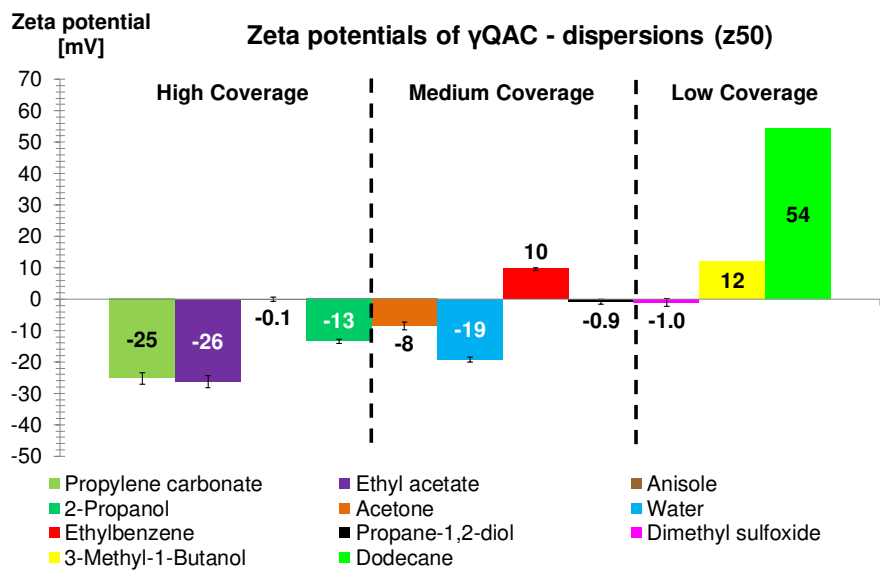

**Figure S16.** Comparison of the median values (z50) of the ζ potential distributions of γQAC dispersions with varying dispersing agents. The error bars indicate twice the standard deviation.

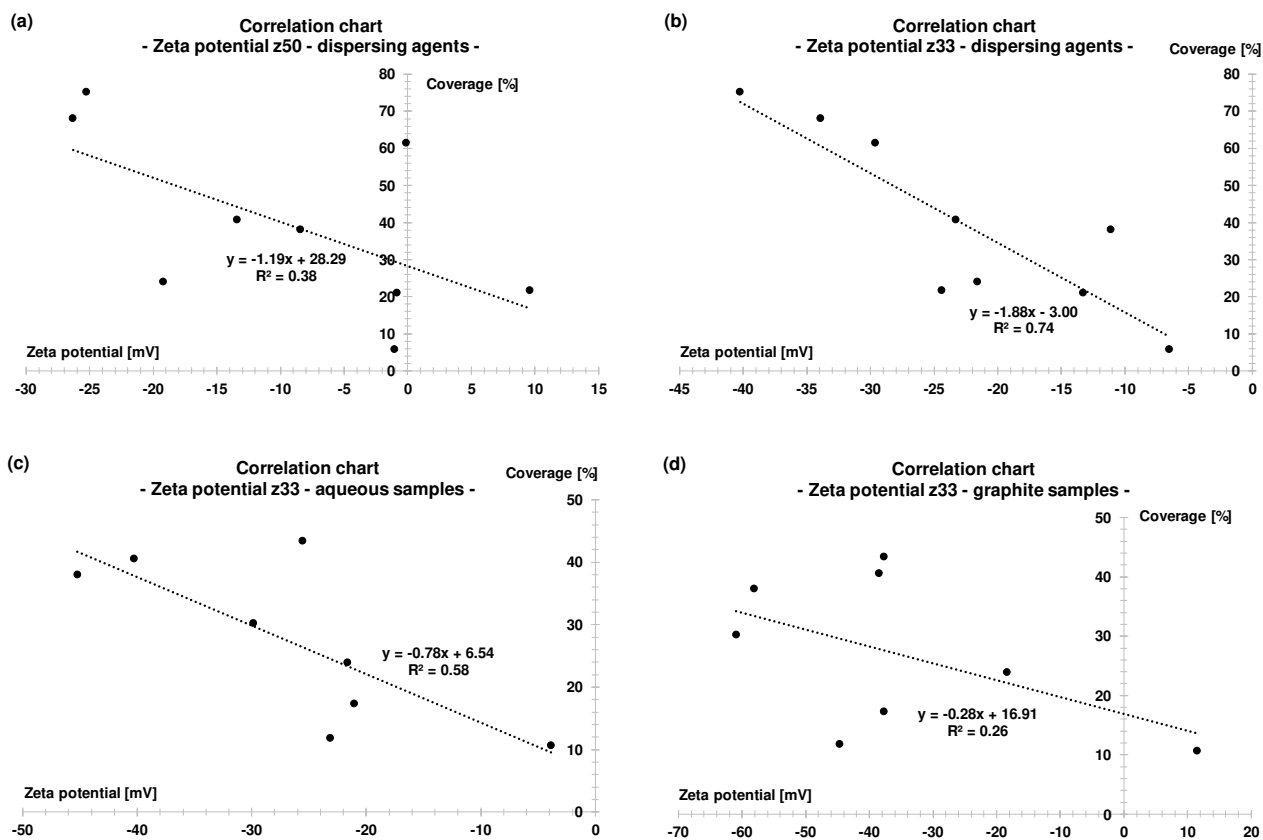

**Figure S17.** Scatter plots relating the median of the surface coverage of samples prepared with different dispersing agents to (a) the median of the ζ potential distribution z50 and (b) the z33 value of the ζ potential distribution. Further, scatter plots relating the median of the achieved surface coverage to (c) the ζ potential z33 of aqueous γQAC dispersions (modified with different salts, KOH, or H₂SO₄) and (d) the ζ potential z33 of aqueous dispersions containing dispersed graphite particles (instead of γQAC).

### 9. The origin of the ζ potential in in aqueous, moderate polar, and non-polar systems.

The dispersing of any object in an aqueous system generates a surface charge and thereby a ζ potential, arising mainly due to the adsorption of ionic species from the liquid phase or by the dissociation of surface molecules.[19,26] The charging mechanism in leaky dielectrics accord with aqueous systems, since the dissociation of ionic compounds is possible.[26-30] However, the Bjerrum length, the distance between ions required for stable dissociation, is significantly higher for fully non-polar media than for aqueous or moderately polar liquids. Further, the occurrence of ionic species is unlikely for the fully non-polar media according to the thermodynamic principles.[26,29,31-33] However, as substantial ζ potentials have long been investigated in non-polar media, it calls for the presence of some other charging mechanism in such media.[26,29-35] A key role in this regards can be attributed to polar impurities, particularly water, that are adsorbed to particle surfaces enabling dissociation processes. Reverse micelles play another important role in this respect: whereas on one hand they help to screen the surface charges and polar adsorbate layers from their non-polar environment, on the other hand, these reverse micelles (and probably other structures) carry charges themselves, obtained via proton transfers, as a result of non-specific thermal interactions. Further, indications were found that the charging of particles that are dispersed in organic liquids is related to molecular charge transfers, being the result of donor-acceptor interactions.[28] The letter results were found by analysing a number of organic liquids, including leaky dielectrics and non-polar media.

## 10. pH measurements

pH measurements were conducted using the pH meter "Checker" from HANNA, allowing a measurement accuracy of $\pm$ 0.2 pH. The sample composition was similar to that of the STM samples: 4 ml sample containing 2 wt. % pigment, and if necessary, additional 0.5 wt. % of salt. Further, prior to each pH measurement, the pH meter was calibrated using either a neutral (7.01), an acidic (pH 4.01), or a basic (pH 10.01) buffer solution, as per the nature of the analyzed sample.

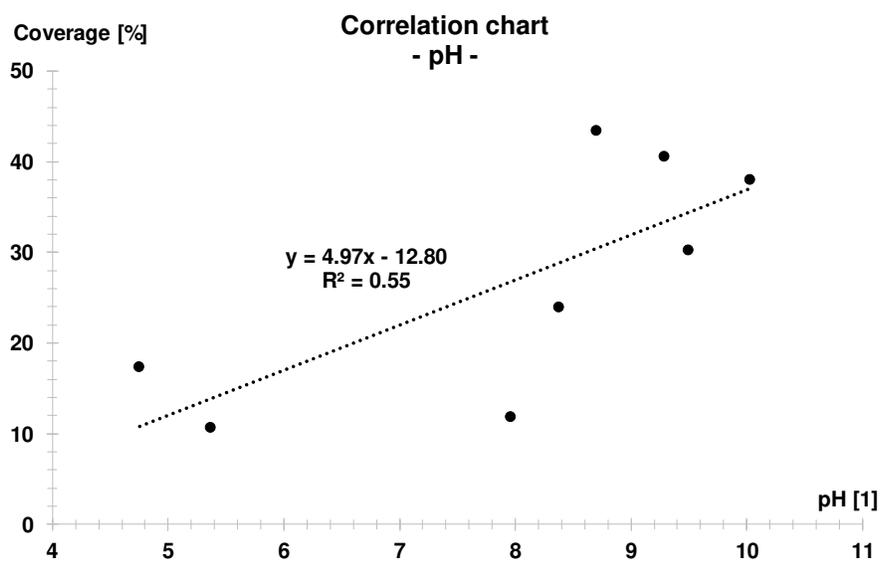

**Figure S18.** Scatter plot relating the median of the achieved surface coverage to the pH.

**11. Scanning electron microscope measurements**

The corresponding measurements have been performed using the scanning electron microscope 440i from Zeiss; the chief scan settings being as: working distance of either 3 or 5 mm and extra-high tension of 20.00 kV. Further, the SE1 detector was used for such measurements.

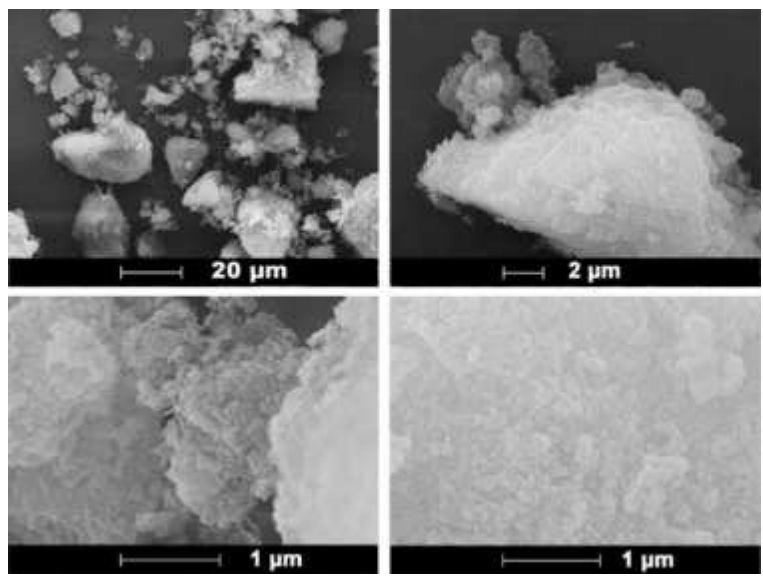

**Figure S19**. Scanning electron microscopy (SEM) pictures of γQAC crystals with increasing magnifications.

## 12. Force field calculations

Classical molecular mechanics calculations have been employed to determine both the binding energy of an organic semiconductor molecule (OSM) within a surface plane of a crystal (comprising of these OSMs) and the adsorption energy of an OSM on a graphite slab (0001). For analysis, three different force fields of the Accelrys' Materials Studio package have been applied, viz. the Dreiding force field, the Universal force field (UFF), and the Consistent Valence force field (CVFF).[13-15] To assign partial charges to the atoms of OSM's atoms, the Gasteiger method, for Dreiding and UFF, has been applied.[16] However, an explicit assignment of a partial charging method is not required for CVFF, since this force field already includes a charging method.[17-18]

Besides, the geometry optimization calculations (3D-periodic) of the crystals' unit cell have been performed according to the so-called "ultra-fine" setting, which corresponds to the convergence threshold values, as shown in the Table S1. However, for all the other calculations comprising larger systems (2D-periodic), due to the huge amount of molecules involved, the so-called "fine" setting has been applied (the terms "fine" and "ultra-fine" refer to settings within Accelrys' Materials Studio).

***Table S1. Convergence threshold values***

|  | **Fine** | **Ultra-fine** |
|---|---|---|
| Energy [kcal / mol] | $1 * 10^{-4}$ | $2 * 10^{-5}$ |
| Force [kcal / (mol Å)] | 0.005 | 0.001 |
| Stress [GPa] | 0.005 | 0.001 |
| Displacement [Å] | $5 * 10^{-5}$ | $1 * 10^{-5}$ |

To determine the adsorption energy of an OSM on graphite, a 2D-periodic system was built with four graphene layers, with the bottom layer fixed. In a distance of 500 Å above these layers, an OSM was set. A geometry optimization calculation was performed to obtain the energy of the relaxed graphite layers plus the energy of an isolated OSM. Next, the OSM was moved on top of the graphite surface (distance about 3.5Å), and the geometry optimization was repeated to obtain the energy of the system where the OSM interacts with the graphite surface. The adsorption energy of the OSM on graphite could then be calculated from the difference between both the energy values.

The binding energy of an OSM located within the surface layer of an OSM crystal was calculated in a similar way. Calculations were performed for the three main crystal plains of the OSM crystal (refer Figure S20 for the crystal planes of γQAC). A small OSM crystal was built by repeating the unit cell of the OSM crystal in the three directions. Further, a 2D-periodic system was created containing the small OSM crystal (periodically repeated in two directions); the bottom crystal layer (of the non-periodic direction) being kept fixed. Depending on the γQAC crystal plane in use for the performed calculations, the nanocrystals were built by: for the crystal plane (100) by 6 x 6 x 3 γQAC unit cells, and for both the crystal planes (010) and (001) by 3 x 6 x 3 γQAC unit cells. Further, an initial geometry optimization calculation was performed to obtain a crystal with a relaxed surface. A second geometry optimization calculation was performed after one OSM of the surface layer had been moved away from the surface by a distance of 500 Å (refer Figure S21). The difference between both the energy values then led to the binding energy of an OSM within a surface layer.

Further, to roughly estimate the error within the binding/adsorption energy calculations, the energies were calculated once again with a system doubled in each of the direction. Such a calculation was performed only for one certain system, and one force field (QAC with Dreiding force field), owing to the long computing times (of several months) for such a largely extended system.

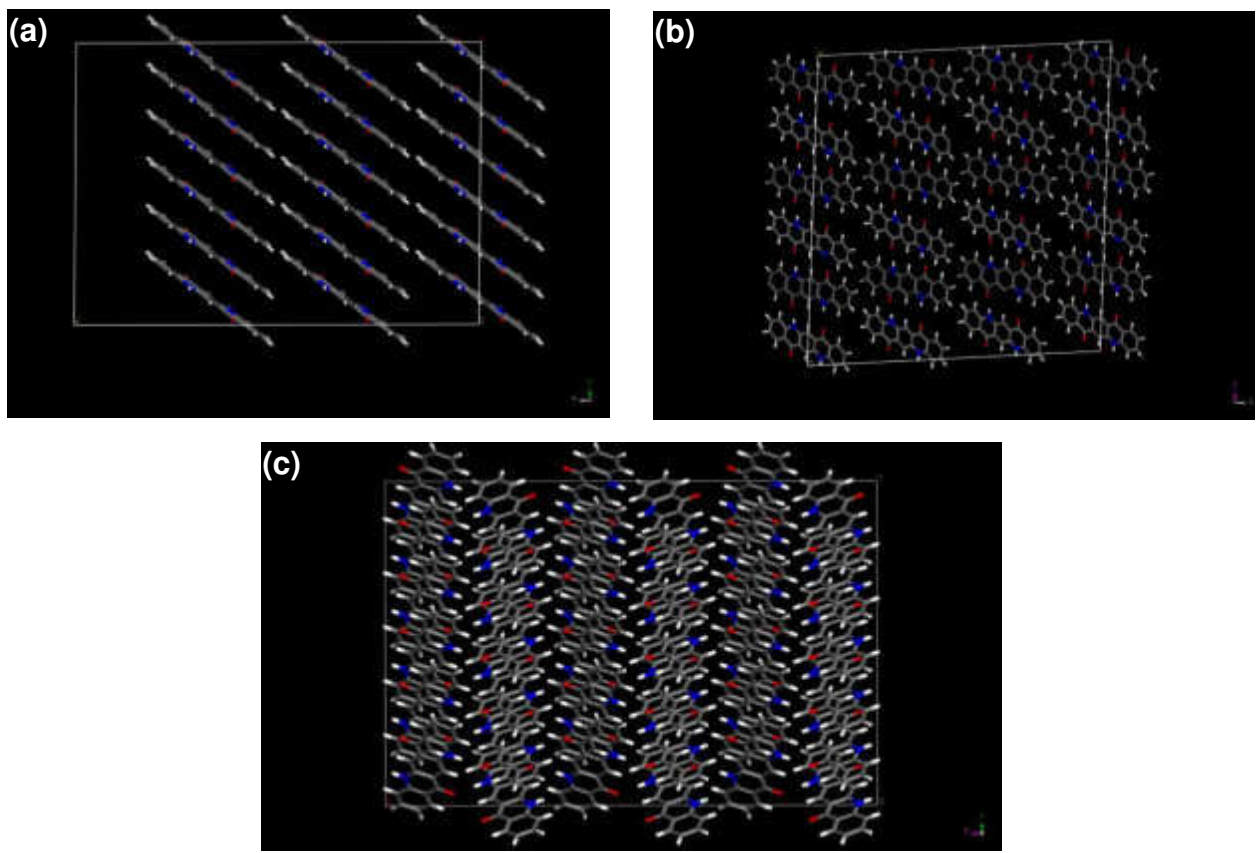

**Figure S20.** Example images depicting the crystal planes of a γQAC nanocrystal, with: (a) the crystal plane (001), (b) the crystal plane (010), and (c) the crystal plane (100).

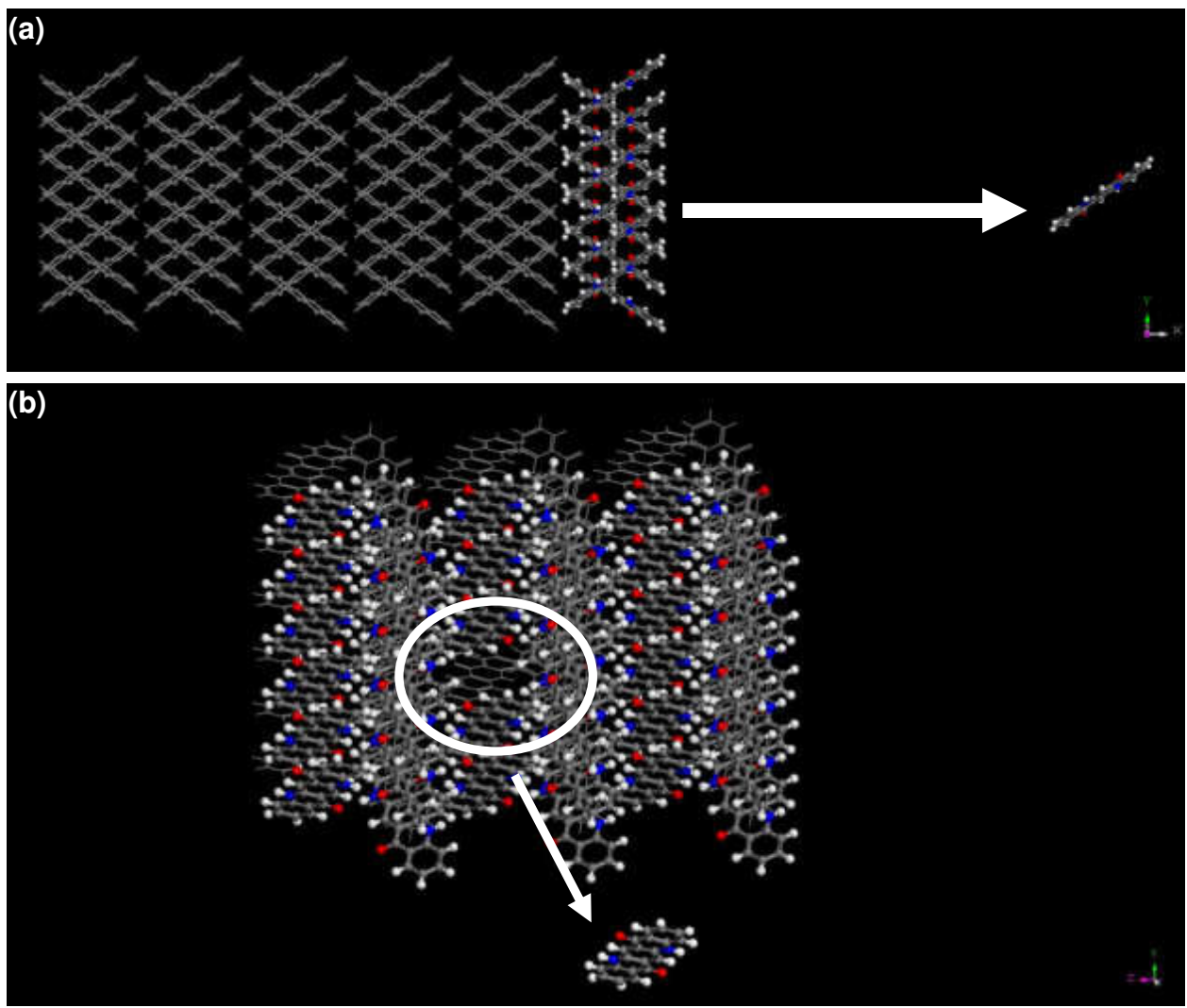

**Figure S21.** Example images depicting how one OSM is moved away from a (100) surface crystal plane of a nanocrystal, in order to determine its binding energy via force field calculations, with: (a) side view and (b) top view. Note: the distance between OSM and nanocrystal depicted in these example images is not true to scale, i.e. it is lower than the one used for the force field calculations (i.e. lower than 500 Å).

## 13. Back calculation determining the γQAC crystal dimensions and a suitable DLVO model

The question had to be assessed whether a γQAC crystal interacts as a whole with the substrate surface during the OSWD process, then for its shape being considered as a spherical shaped particle (Figure S19). However, powdered γQAC, as commercial semiconductors in general, does not comprise of mono-crystals, but agglomerated particles having very uneven resulting surfaces (Figure S19). Referring to the limited operating range of the π-π interactions, a γQAC agglomerate might therefore just partially interact with the HOPG surface, hence the interacting section, potentially, being considered as plate-like for such a case. To explore the above, the following approach was pursued: based on the number of QAC molecules, constituting well-defined, two-dimensional supramolecular arrays, the idea was to determine the dimensions of the source-crystals and compare them with the actual measured particle sizes.

The force field calculations discussed in the article text and the limited operating range of π-π interactions led to the assumption that a cohesive (010) crystal face delivers the QAC molecules to form a single cohesive, two-dimensional array, via the OSWD. Based on this assumption, a rough calculation to determine the equivalent diameter of a single γQAC crystal, having a (010) crystal face with the desired amount of QAC molecules, was performed. Since each unit cell of a γQAC crystal lattice contains 2 QAC molecules,[20] the quantity of QAC molecules $N_M$ in the crystal face (010) can thus be given by the relation (using the variable $n = \{1; 2; 3; 4; \ldots\}$:

$$N_M(n) = 2\,n^2 \tag{4}$$

Using the above equation, the volume of the unit cell $V_{UC}$,[20] and the given quantity of QAC molecules $N_M$, the volume of the original γQAC crystal $V_{QAC}$ can be related as:

$$V_{QAC} = \left(\frac{N_M}{2}\right)^{\frac{3}{2}} * V_{UC} \tag{5}$$

The equivalent diameter $D_e$ of the γQAC crystal can hence be calculated by the formula:

$$D_e = 2 * \sqrt[3]{\frac{V_{QAC}}{\pi} * \frac{3}{4}}$$ (6)

Example values for these variables can be found in Table S2, showing exemplary the results of these back calculations.

In this regard, the number of the QAC molecules constituting an array was found to be in the range of 30 – 2655 molecules. Hence, the results of the related back calculations revealed that the equivalent diameters of the detaching γQAC crystals were in the range of 4 – 41 nm. Comparing these results with the measured values of 5.6 – 28.4 µm (Figure S14 (a)), it was hence deduced that just a tiny part of a γQAC crystal (approx. 0.3 – 1 ppm of its surface) interacts with the HOPG surface. Consequently, it was concluded that the interactions among the semiconductor crystal and the underlying substrate surface can be best described by the DLVO model for the plate-like interactions.[19]

**Table S2. Example values of the variables used in the performed back calulations**

| $n$ [1][a] | $N_M(n)$ [1][b] | $V_{QAC}$ [nm³][c] | $D_e$ [nm][d] |
|---|---|---|---|
| 4 | 32 | 44 | 4 |
| 10 | 200 | 691 | 11 |
| 15 | 450 | 2334 | 16 |
| 20 | 800 | 5532 | 22 |
| 25 | 1250 | 10804 | 27 |
| 30 | 1800 | 18669 | 33 |
| 34 | 2312 | 27177 | 37 |
| 37 | 2738 | 35024 | 41 |

[a]The variable $n$. [b]The dedicated quantity of QAC molecules within the crystal face (010) $N_M$. [c]The volume of the related γQAC crystal $V_{QAC}$. [d]The thus calculated equivalent diameter $D_e$ of the γQAC crystal.

## 14. Traditional DLVO simulations – relevant equations

The potential energy of interaction $V_T(h)$ as a function of the separation $h$ between the HOPG and the γQAC surface was determined as per the DLVO theory for parallel plates,[19] as:

$$V_T(h) = \left(-\frac{H_{123}}{12\,\pi} * \frac{1}{h^2}\right) + \left(\frac{64\,c_{i0}\,k\,T\,\Gamma_1\,\Gamma_3\,e^{(-\kappa h)}}{\kappa} * \frac{1000\,N_{Av}}{1}\right) \tag{7}$$

Where $H_{123}$ is the Hamaker constant for the system: HOPG (index 1), dispersing agent (index 2), and the γQAC (index 3), given as:

$$H_{123} = \left(\sqrt{H_{11}} - \sqrt{H_{22}}\right) * \left(\sqrt{H_{33}} - \sqrt{H_{22}}\right) \tag{8}$$

Further, the Hamaker constant $H_{ii}$ for the individual phases can be written as:

$$H_{ii} = \left(\frac{\rho\,N_{Av}\,\pi}{M}\right)^2 * \frac{3}{4} * \frac{\alpha^2\,ℏ\nu}{(4\,\pi\,\varepsilon_0)^2} \tag{9}$$

The polarizability $\alpha$ in the above equation, being determined via the Clausius-Mosotti relation,[21] as:

$$\alpha = \left(\frac{\varepsilon_r - 1}{\varepsilon_r + 2}\right) * \frac{M}{\rho} * \frac{3\,\varepsilon_0}{N_A} \tag{10}$$

The reverse Debye screening length $\kappa$ was calculated by the relation:

$$\kappa = \sqrt{\frac{1000\,e_0^2\,N_{Av}\,\sum z_i^2\,c_{i0}}{\varepsilon_r\,\varepsilon_0\,k\,T}} \tag{11}$$

And the Goy-Chapman coefficients $\Gamma_i$ calculated by:

$$\Gamma_i = \frac{|z_i|\,e_0\,\Phi_i}{4\,k\,T} \tag{12}$$

The constants used in these equations being: the molar electrolyte concentration $c_{i0}$, the Boltzmann's constant $k$, the temperature $T$ (293,15 K), the Avogadro's number $N_{Av}$, the density $\rho$, the molar mass $M$, the ionization energy $ℏ\nu$,[23-25] the vacuum permittivity $\varepsilon_0$, the relative permittivity $\varepsilon_r$, the valence including the sign $z_i$, the electronic charge $e_0$, and the surface potential (the $\zeta$ potential) $\Phi_i$.

### 15. Traditional DLVO simulations – results

In the course of the presented study, additional simulations were performed utilizing the traditional DLVO theory. The appropriate results are described below. To start with, calculations were done using all the positively charged ions present within the different aqueous systems: $H_3O^+$ being always present, the phosphate salts and sodium chloride delivering $Na^+$, KOH delivering $K^+$. The results of these calculations being demonstrated in the Figure S22; as can be seen, the single interaction energy functions exhibit strongly varying trends in the positive range, specifying the repulsive energy barrier (Figure S22, (a)). A comparison of the energy barrier maxima (Figure S22, (b)) reveals a trend towards aqueous systems catalyzing a high substrate surface coverage to exhibit a distinct repulsive energy barrier, whereas the systems water + sodium chloride, water + disodium pyrophosphate, and water + $H_2SO_4$ do not feature any repulsive energy barrier. However, such result being not conclusive, since a distinct repulsive energy barrier prevents a physical contact between substrate and dispersed semiconductor crystal, what was found to be a basic requirement for both the OSWD and the solid-solid wetting effects in general.[36-40]

Referring to the explanations in the main article, further DLVO calculations were performed by solely considering the concentration of the $H_3O^+$ - ions, taken from the previously measured pH values. Results revealed a distinctly lesser level and a flattened trend of the interaction energy functions (Figure S23, (a)), with only the sample modified with disodium phosphate displaying a narrow and tall repulsive energy peak. Further, the comparison of the energy barrier maxima yielded the water + $H_2SO_4$ sample to not feature any repulsive energy barrier and such result being still not conclusive (Figure S23, (b)).

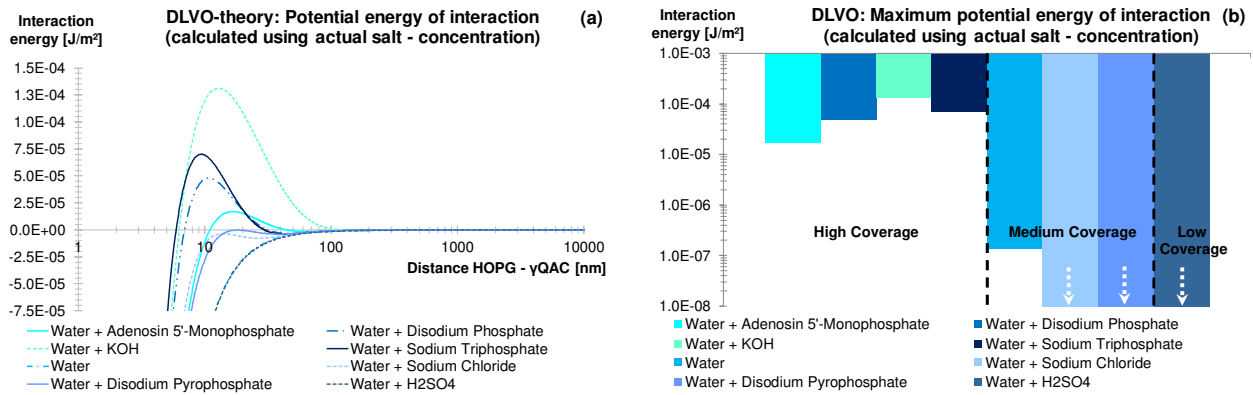

**Figure S22.** Simulation of the potential energy of interaction for the aqueous systems. (a) The complete interaction energy functions. (b) Comparison of the interaction energy maxima. Note: the samples water + sodium chloride, water + disodium pyrophosphate, and water + $H_2SO_4$ do not feature a repulsive energy barrier.

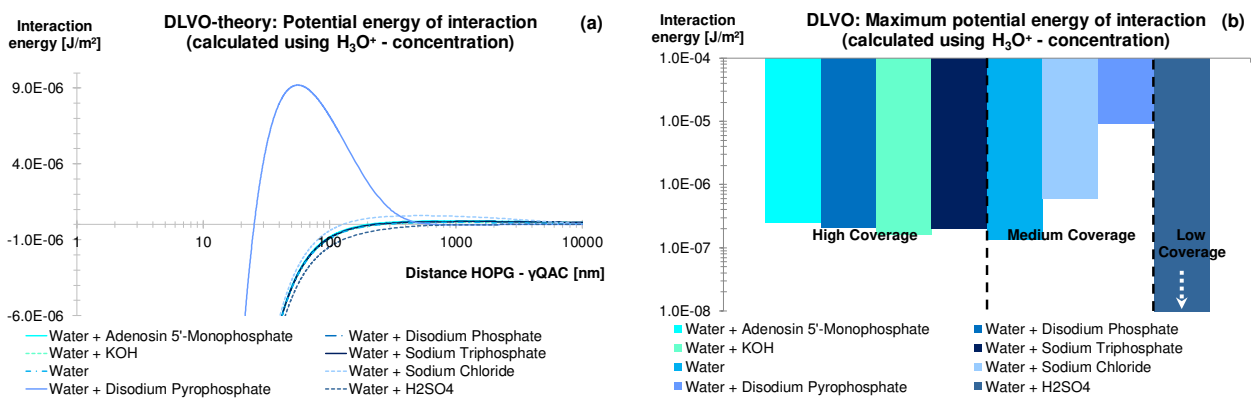

**Figure S23.** DLVO simulations by using solely the concentration of $H_3O^+$ - ions to determine the repulsive double layer forces. (a) The complete interaction energy functions. (b) Comparison of the interaction energy maxima. Note, that the sample water + $H_2SO_4$ does not feature a repulsive energy barrier.

## 16. Refined DLVO simulations – relevant equations

As is described in the article, the traditional DLVO simulations performed during this study were adjusted, in accordance with the work of Lukatsky and Safran, so as to include all kinds of fluctuation-induces forces.[22] The calculations are valid for particle separations that are significantly smaller than the Goy-Chapman length, which is approx. 10 µm in our aqueous systems. This enhanced term for the calculation of the potential energy of interaction $V_{eT}$ consists of three sub-parts: the attractive Van-der-Waals interaction (calculated again via the Hamaker theory $V_H$), a new term for the unscreened Poisson-Boltzmann repulsion $V_{PB}$, and an additional term for the fluctuation-induced forces $V_{ff}$, expressed as:

$$V_{eT}(h) = V_H + V_{PB} + V_{ff} \tag{13}$$

$$V_H(h) = -\frac{H_{123}}{12\,\pi} * \frac{1}{h^2} \tag{14}$$

Note, that the originally determined interaction pressures $\Pi_n$ by Lukatsky and Safran, were by us converted into the interaction energies $V_n$,[19] via the relation:

$$V_n(h) = -\int \Pi_n dh \tag{15}$$

As a result, the Poisson-Boltzmann repulsion $V_{PB}$ and the fluctuation-induced forces $V_{ff}$, are related as:

$$V_{PB}(h) = -\frac{k\,T}{\pi\,\lambda\,l_B}\ln(h) \tag{16}$$

$$V_{ff} = \frac{7\,k\,T}{4\,\pi\,\lambda^2}\ln(h) \tag{17}$$

The Gouy-Chapman length $\lambda$ is calculated by:

$$\lambda = \frac{1}{2\,\pi\,l_B\,\sigma_0} \tag{18}$$

The Bjerrum length $l_B$ is calculated by:

$$l_B = \frac{e_0^2}{4\,\pi\,\varepsilon_2\,\varepsilon_0\,k\,T} \tag{19}$$

The total surface charge number density $\sigma_0$ is calculated by: [19]

$$\sigma_0 = \frac{1}{2}\left(\frac{\varepsilon_2\,\varepsilon_0\,\kappa\,(\Phi_1 + \Phi_3)}{e_0}\right) \tag{20}$$

### 17. Refined DLVO simulations – calculations using the actual salt-concentration

Via the refined DLVO theory, additional simulations were performed using all the positively charged ions present within the different aqueous systems (Figure S24). Analyzing the calculated interaction energy functions, results revealed one significant repulsive energy barrier related to the sample water + KOH. As stated before, a distinct repulsive energy barrier would significantly limit the potential of triggering the OSWD, making the latter result not conclusive as it does not match the experimental results that were obtained in this regard.

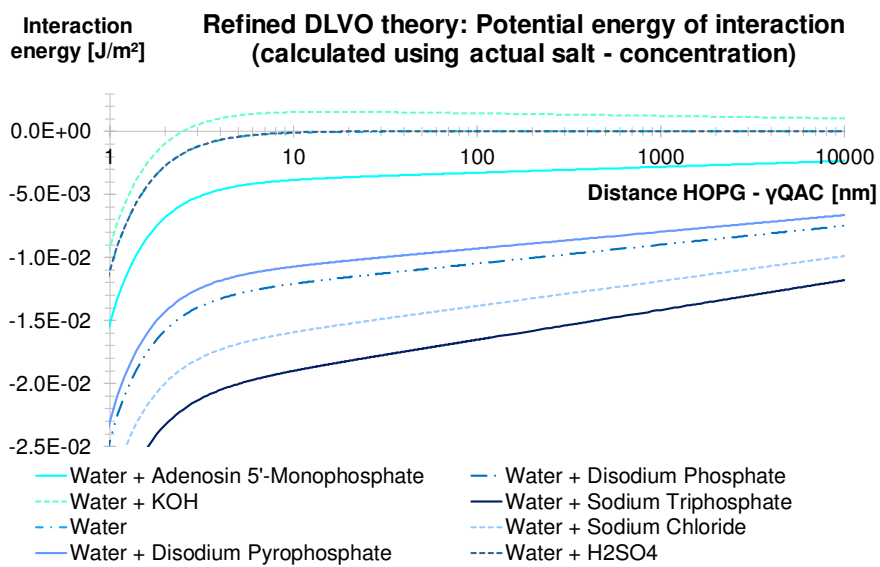

**Figure S24.** Interaction energy functions, determined via refined DLVO simulations using all the positively charged ions present within the different aqueous systems.

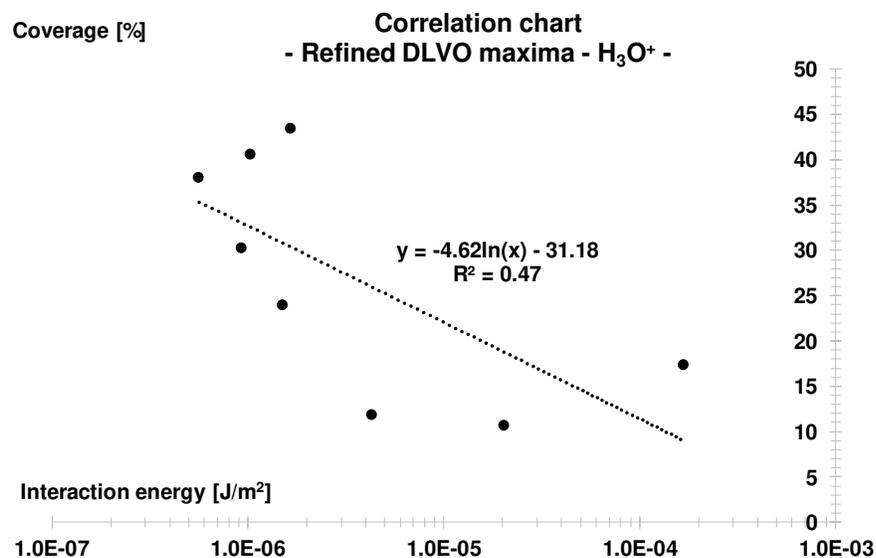

**Figure S25.** Scatter plot relating the median of the achieved surface coverage to the repulsive energy barrier, determined via refined DLVO calculations (considering solely the concentration of $H_3O^+$ ions).

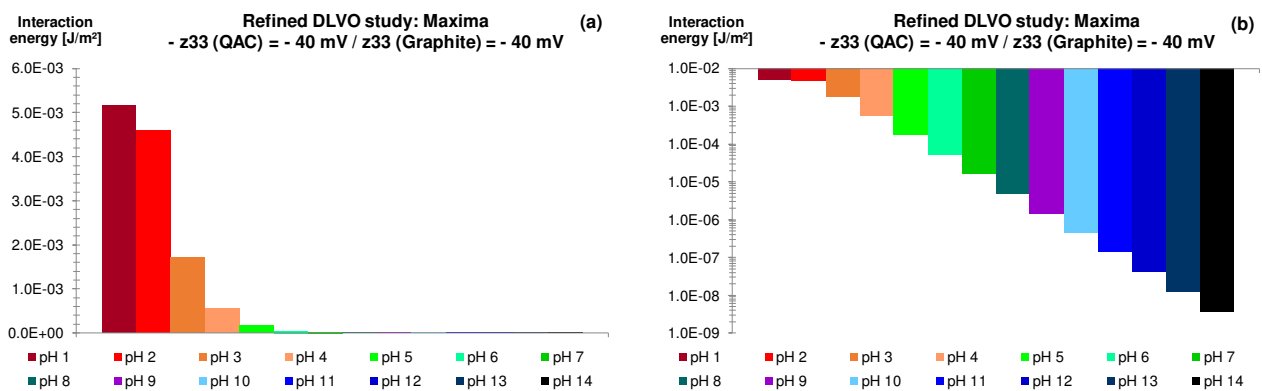

**Figure S26.** Results of a comparative study, calculating the interaction energy maximum for different pH values, with all other conditions being kept identical ($z33$ of QAC = $z33$ of Graphite = - 40 mV; Temperature = 293.15 K). (a) Diagram showing the calculated DLVO maxima. The DLVO maximum decreases with increasing pH value. (b) Diagram showing the results of the same study, but with a logarithmical scale of the y-axis.